\documentclass[STIX1COL]{WileyNJD-v2}
\usepackage{moreverb}

\newcommand\BibTeX{{\rmfamily B\kern-.05em \textsc{i\kern-.025em b}\kern-.08em
T\kern-.1667em\lower.7ex\hbox{E}\kern-.125emX}}

\articletype{Original Paper}%

\received{<day> <Month>, <year>}
\revised{<day> <Month>, <year>}
\accepted{<day> <Month>, <year>}

\usepackage{commath}
\usepackage[numbers,sort&compress,comma,square]{NJDnatbib}
\renewcommand\thefigure{\arabic{figure}}
\renewcommand\thetable{\arabic{table}}

\usepackage{amsmath,amsthm,bm}
\usepackage{color}
\usepackage{graphicx}
\usepackage{subfig}

\newcommand{\rd}{\mathrm{d}}
\newcommand{\re}{\mathrm{e}}

\usepackage{hyperref}
\hypersetup{
    colorlinks=true,
    linkcolor=blue,
    filecolor=magenta,      
    urlcolor=cyan,
    breaklinks=true
}

\begin{document}

\title{Revisiting steady viscous flow of a generalized Newtonian fluid through a slender elastic tube using shell theory}

\author[]{Vishal Anand}
\author[]{Ivan C.\ Christov*}

\address[]{School of Mechanical Engineering, Purdue University, West Lafayette, Indiana 47907, USA}

\corres{*Ivan C.\ Christov, School of Mechanical Engineering, Purdue University, West Lafayette, Indiana 47907, USA. \email{christov@purdue.edu}}

\abstract[Abstract]{
A {flow vessel with an elastic wall} can deform significantly due to viscous fluid flow within it, even at vanishing Reynolds number (no fluid inertia). Deformation leads to an enhancement of throughput due to the change in cross-sectional area. The latter gives rise to a non-constant pressure gradient in the flow-wise direction and, hence, to a nonlinear flow rate--pressure drop relation (unlike the Hagen--Poiseuille law for a rigid tube). Many biofluids are non-Newtonian, and are well approximated by generalized Newtonian (say, power-law) rheological models. Consequently, we analyze the problem of steady low Reynolds number flow of a generalized Newtonian fluid through a slender elastic tube by coupling fluid lubrication theory to a structural problem posed in terms of Donnell shell theory. A perturbative approach (in the slenderness parameter) yields analytical solutions for both the flow and the deformation. Using matched asymptotics, we obtain a uniformly valid solution for the tube's radial displacement, which features both a boundary layer and a corner layer caused by localized bending near the clamped ends. In doing so, we obtain a ``generalized Hagen--Poiseuille law'' for soft microtubes. We benchmark the mathematical predictions against three-dimensional two-way coupled direct numerical simulations (DNS) of flow and deformation performed using the commercial computational engineering platform by ANSYS. The simulations  show good agreement and establish the range of validity of the theory. Finally, we discuss the implications of the theory on the problem of the flow-induced deformation of a blood vessel, which is featured in some textbooks.
}


\jnlcitation{\cname{%
\author{V.~Anand} and  
\author{I.~C.~Christov} 
(\cyear{2020}), 
\ctitle{Revisiting steady viscous flow of a generalized Newtonian fluid through a slender elastic tube using shell theory}, \cjournal{Z Angew Math Mech}, \cvol{2020}, accepted for publication.}}

\maketitle


\section{Introduction}

Microfluidics is the study of the manipulation of microscopic volumes of fluids at small scales \citep{SQ05}. Though the subfield of fluid mechanics pertaining to flows at small scales, i.e., \emph{low Reynolds number hydrodynamics} \citep{HB83}, is not a new field, its relevance to technologies at the microscale (and thus the emergence of the term ``microfluidics'') was realized only in the 1990s \citep{B08}. Technological advancements in microfabrication over the past few decades \citep{XW98,KRSDC12,SCFT16} have made microscale fluid mechanics more accessible experimentally. A number of applications have emerged, including the connection between microfluidics and technologies for global medial and social problems \citep{W06}, such as lab-on-a-CD medical diagnostics \citep{Madou2006,Kong2016} and platforms fot \textit{in vitro} isolation of cancer cells \citep{Nagrath2007}, amongst a variety of lab-on-a-chip devices \citep{Abgrall2007} and micro-total analysis ($\mu$TAS) systems \citep{RDIAM02,AIRM02}. Microfluidic devices afford many advantages over their traditional counterparts: portability, low reagent consumption and short analyses times, often at higher resolutions than macroscopic counterparts \citep{NW06}.

A feature of microscale fluid mechanics is that fluid--structure interactions (FSIs) occur in both external and confined flows due to the compliance of the various solid-wall materials \citep{DS16,KCC18}. For example, in inertialess locomotion, a swimmer may be deformable (e.g., a bacterium has a flexible flagellum) or the fluid may be ``deformable'' (e.g., a polymeric substance dissolved into a liquid). On the one hand, in external flows, the elasticity of a swimmer affects its propulsive thrust and ability to navigate \citep{Lauga16}. On the other hand, in confined flows, the flow conduit may be made of a deformable material \citep{KCC18}, such as polydimethylsiloxane (PDMS) (a polymeric gel) \citep{XW98} or elastin (a constituent of arteries) \citep{SGF77}. Then, ``creeping'' flows can delaminate an elastic membrane from a solid boundary, forming blisters \citep{CVB08}, which are prone to a wealth of mechanical instabilities \citep{JPPH18} and whose inflation dynamics are sensitive to the contact line conditions \citep{HBDB14}. The hydrodynamic pressure within such conduits is affected by their deformation due to two-way FSI. Extensive experimental work over the past decade has sought to  understand FSI in microfluidics, specifically the effect of FSI on the flow rate--pressure drop relationship in a soft microchannel  \citep{GEGJ06,SLHLUB09,OYE13,MY19}. Specifically, {for the case of steady, low Reynolds number flow in microchannels for which the ambient and outlet pressures are the same (no extramural pressure differences)}, the pressure drop across a soft microchannel is significantly smaller compared to the rigid case. Consequently, deviations are expected from the classical \emph{Hagen--Poiseuille law} \citep{SS93}, which relates the viscous pressure drop across a length of pipe to the volumetric flow rate through it, the fluid properties and the pipe dimensions. A goal of the present study of \emph{microscale fluid--structure interactions} is to mathematically analyze and quantify such deviations for axisymmetric geometries (i.e., microtubes rather than microchannels).

The study of moderate Reynolds number instabilities due to FSIs in elastic tubes is a time-honored subject \citep{P80,G94,GJ04,HH11,KHG12}, primarily due its relevance to biofluid mechanics of the arteries and the lungs, for example in the contraction of the trachea during coughing \citep{G94}. Consequently, microtubes traditionally reside on the opposite end of the FSI spectrum from microchannels, in terms of Reynolds number. Specifically, the study of \emph{collapsible tubes} concerns conduits whose radius \textit{decreases} owing to a negative transmural pressure difference, eventually completely collapsing \citep{S77,BP82,BT06}. 
Here, the flow field is approximated as one-dimensional, by averaging across the cross-section, but it is not fully-developed \citep{S77}. Viscous effects are captured using a pipe flow friction factor. A \emph{tube law} is obtained to relate the local transmural pressure difference to the change in area due to circumferential and axial bending and tension, from postulated simple relations \citep{S77} to rigorous derivations from shell theory \citep{WHJW10}. The mathematical analysis of stability of such three-dimensional (3D) flows is an ongoing challenge \citep{GBGP18}. 

Nevertheless, there is also a need to develop accurate models for low Reynolds Number FSI in soft tubes due to the relevance to blood flow through small arteries \citep{CM03}. Meng et al.~\citep{MHT15} studied such a problem of drainage of liquids from collapsible tubes of circular and elliptical cross-sections. The tubes had been stretched initially with a prescribed tension in the radial direction and then filled with a fluid. The pressure built up inside the fluid is then a function of the prescribed tension. Assuming plane-strain conditions inside the structure with zero net axial force and the Hagen--Poiseuille law for the Newtonian flow within, Meng et al.~\citep{MHT15} obtained a set of differential equations governing the evolution of the tube axes. Neither the flow rate nor pressure drop across the tube were specified initially, but they could be computed by this approach.

Most of the latter research has focused on Newtonian fluids. Biofluids are, however, non-Newtonian \citep{C05,Lee06}. Blood is often modeled {as a Casson fluid}, which has both a yield stress and a shear-dependent viscosity \citep[Ch.~3]{F93}. Research on microscale FSIs has only just begun to take into account the non-Newtonian nature of the working fluids \citep{RS16,BBG17,RCDC18,ADC18}. Raj M et al.~\citep{RCDC18} performed experiments on FSIs in a microchannel with \emph{circular} cross-section, which is more akin to a  blood vessel. This conduit was fabricated from PDMS using a pull-out soft lithography process form a large slab. Xanthan gum was used as a non-Newtonian blood-analog fluid. Measurements of the pressure drop at different inlet flow rates were shown to match a simple mathematical model of one-way FSI, meaning that the pressure was calculated using the Hagen--Poiseuille law for a rigid tube and then imposed as a load on the structure, without coupling the microchannel shape change back into the hydrodynamics. Microtubes and microchannels of circular cross-section are now of significant scientific interest due to ``a new, direct peeling-based technique for building long and thin, highly deformable microtubes'' \citep{PCK15}, which can be used in building bioinspired and biocompatible soft devices \citep{Wu18}.

In a recent series of works \citep{EG14,EG16,BBG17}, two-way FSI coupling was employed to analyze the transient pressure and deformation characteristics of a shallow, deformable microtube. Employing the Love--Kirchhoff hypothesis, a relation was obtained between the internal pressure load in a soft tube and its radial and axial deformations, up to the leading order in slenderness. Treating the structural problem as quasi-static, an unsteady diffusion-like equation for the fluid pressure was obtained and analyzed for both Newtonian \citep{EG14,EG16} and generalized Newtonian \citep{BBG17} fluids. However, the effect of non-trivial deformation on the resulting flow rate--pressure drop relation for the tube (transient or steady-state) was not considered or benchmarked against simulations and/or experiments. Anand et al.~\citep{ADC18} discussed the former for FSI between a microchannel and a generalized Newtonian fluid. Meanwhile, Vedeneev et al.~\citep{VP18} obtained results on moderate-Reynolds-number instabilities in collapsible tubes conveying generalized Newtonian fluids. However, the steady problem has not received much attention, and a detailed study of the latter is the goal of this work.

Thus, it should be clear that microscale fluid--structure interactions (FSIs) are an important problem in mechanics, and the recent literature {suggests} that asymptotic and perturbation methods can be used to develop theories of these FSIs. To this end, in this paper, we present a comprehensive theoretical and computational study of steady non-Newtonian FSIs in deformable tubes. To account for the non-Newtonian rheology of biofluids, we employ a generalized Newtonian model with a power-law shear-dependent viscosity, which is suitable for steady flows. To account for the compliance of the (initially) cylindrical flow conduit, we employ classical (linear) shell theories. In \S\ref{section:prelims}, we describe the problem of interest, specifying the physical domain and the notation. Then, in \S\ref{sec:flow}, we solve for the flow field under the lubrication approximation. In \S\ref{section:deformation}, we employ thin-shell theory to solve for the deformation field. We bring all this together in \S\ref{section:coupling}, wherein the flow and the deformation fields are coupled, yielding a complete theory of steady non-Newtonian FSI in a tube. In \S\ref{section:results}, the predictions of the proposed theory developed are benchmarked against full two-way coupled, 3D direct numerical simulations (DNS) of steady-state FSI carried out using commercial computational engineering tools. The benchmark against DNS allows us to both validate our mathematical results as well as to determine the theory's range of applicability. Finally, \S\ref{section:conclusion} summarizes our results and presents an outlook for future work. To make the present work self-contained, appendices are provided discussing the DNS approach (Appendix~\ref{Appendix:ANSYS}) and the relation of our work to textbook models of flow in elastic blood vessels (Appendix~\ref{Appendix:Fung}).

\section{Preliminaries}
\label{section:prelims}

\begin{figure}
  \centering
  \includegraphics[width=0.75\linewidth]{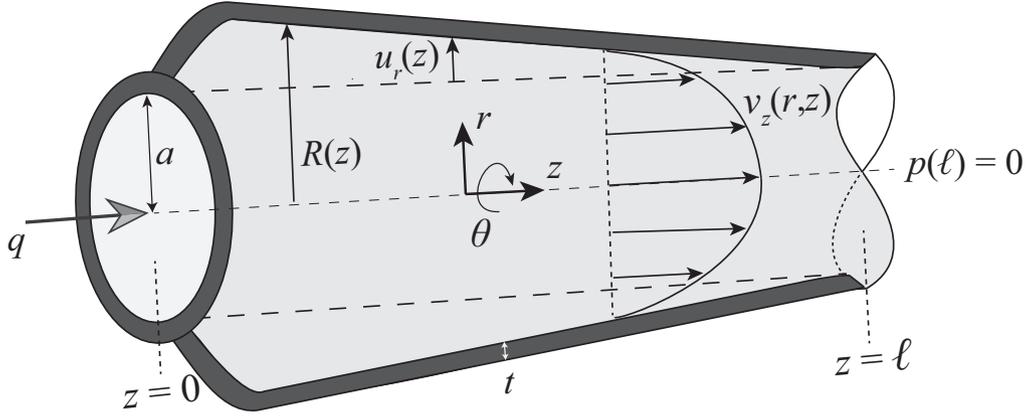}
  \caption{~Schematic of the slender and thin tube geometry in its deformed configuration. The notation for the flow and the deformation is also labeled. [Reprinted/adapted with permission from ``On the Deformation of a Hyperelastic Tube Due to Steady Viscous Flow Within,'' Vishal Anand, Ivan C. Christov, {\it Dynamical Processes in Generalized Continua and Structures}, Advanced Structured Materials {\bf 103}, pp.~17--35, doi:10.1007/978-3-030-11665-1\_2.  \textcopyright\ Springer Nature 2019].}
\label{Figure_MC}
\end{figure}

We consider an initially cylindrical flow conduit geometry. As shown in Fig.~\ref{Figure_MC}, the geometry of the tube is \emph{slender}, i.e., its streamwise dimension is much larger than its cross-sectional dimension, and \emph{shallow} (or, \emph{thin}), i.e., its wall thickness is much smaller than its cross-sectional dimension. The cylindrical coordinate system has its origin at the center of the inlet of the tube but it is displaced in Fig.~\ref{Figure_MC} for clarity. The wall of the tube has a finite thickness, and it is soft; hence, it deforms elastically due to the fluid flow within it. Specifically, the tube has an undeformed radius $a$, a constant length $\ell$, and constant (within the classical shell theories to be discussed below) thickness $t$. The radial deformation of the tube is denoted by $u_r(z)$ so that the radius of the deformed tube is $R(z)=a+u_r(z)$. Axisymmetry ensures that the latter only depends on the streamwise coordinate $z$. The tube's wall is composed of a linearly elastic material with constant modulus of elasticity (Young's modulus) $E$ and a constant Poisson ratio $\nu$.

Fully-developed steady flow of a non-Newtonian fluid enters the tube across the inlet ($z=0$ plane) at a constant flow rate $q$. The non-Newtonian behavior of the fluid is due to its shear-dependent viscosity. Our main objective is to determine the relation between the pressure drop $\Delta p$, across the length of the tube, and the imposed flow rate $q$. In other words, we seek to derive, mathematically, the Hagen--Poiseuille law for steady non-Newtonian flow in a deformable tube. To this end, we simplify the fluid flow (\S\ref{sec:flow}) and structural mechanics (\S\ref{section:deformation}) problems independently in the appropriate asymptotic limit(s). Then, we solve the two sets of governing equations, which are coupled together by the hydrodynamic pressure (the normal forces exerted by the fluid), which act as a load on the structure. 

\section{Fluid mechanics problem}
\label{sec:flow}

The assumptions made pertaining to the fluid flow problem are:
\begin{enumerate}
\item Steady flow: ${\partial}(\,\cdot\,)/{\partial t} = 0$.

\item Axisymmetric flow {without swirl}: ${\partial}(\,\cdot\,)/{\partial \theta} = 0$ and $v_{\theta} = 0$.

\item Slender tube: $\ell \gg a \;\Leftrightarrow\; \lambda := {a}/{\ell} \ll 1$.
\end{enumerate}
Assumption {3} is key to our analysis. Davis~\citep{D17b} highlights the ``importance of being thin'' in making analytical progress on nonlinear fluid mechanics problems.
  
First, we determine the kinematics of the flow. In the cylindrical coordinates labeled in Fig.~\ref{Figure_MC}, and under assumption {2} above, the fluid's equation of continuity (conservation of mass) is
\begin{equation}
  \frac{1}{r}\frac{\partial}{\partial r}(r v_r) + \frac{\partial v_z}{\partial z} = 0.
  \label{eq: cont_dim_reduced}
\end{equation}
Let us now introduce the following dimensionless variables:
\begin{equation}
   \bar{r} = r/a, \qquad
   \bar{z} = z/\ell, \qquad
   \bar{v}_{\bar{r}} = v_r/\mathcal{V}_r, \qquad
   \bar{v}_{\bar{z}} = v_z/\mathcal{V}_z, \qquad
   \bar{p} = p/\mathcal{P}_c, 
\label{eq:nd_vars_tube}
\end{equation}
Here, $\mathcal{V}_z$ and $\mathcal{V}_r$ are characteristic velocity scales in the axial and radial directions respectively, while $\mathcal{P}_c$ is the characteristic pressure (stress) scale: e.g., the full pressure drop in pressure-controlled scenarios or the viscous pressure scale in flow-rate-controlled situations. $\mathcal{P}_c$ can be determined from the velocity scale [see Eq.~\eqref{eq:v_p_scale_rel} below]. 
Introducing the dimensionless variables from Eq.~\eqref{eq:nd_vars_tube}, Eq.~\eqref{eq: cont_dim_reduced} becomes
\begin{equation}
\frac{\mathcal{V}_r}{a} \frac{1}{\bar{r}}\frac{\partial}{\partial \bar{r}}(\bar{r} \bar{v}_{\bar{r}}) + \frac{\mathcal{V}_z}{\ell}\frac{\partial \bar{v}_{\bar{z}}}{\partial \bar{z}} = 0.
\label{eq:nd_cont}
\end{equation}
Balancing all terms in Eq.~\eqref{eq:nd_cont} yields the characteristic radial velocity scale:
$\mathcal{V}_r \equiv  \lambda \mathcal{V}_z$. Consequently, to the leading order in $\lambda$, the velocity field is \emph{unidirectional}: $\bar{\boldsymbol{v}} = \bar{v}_{\bar{z}}(\bar{r}) \boldsymbol{\hat{k}}$, where $\boldsymbol{\hat{k}}$ is the unit normal vector in the $z$-direction. Below, we show that, due to FSI, the unidirectional profile ``picks up'' a weak $\bar{z}$ dependence as well, which is ``allowed'' under the \emph{lubrication approximation} \citep{B08}.
   
Next, we consider the dynamics of the flow field. Since we are dealing with flow at the microscale, the Reynolds number $Re$ (to be properly defined below upon introducing the fluid's rheology) is assumed to be small (i.e., $Re\ll1$), and the lubrication approximation applies. Consequently, inertial forces in fluid are negligible in comparison to pressure and viscous forces, and we begin our analysis with the following simplified equations expressing the momentum conservation in the (radial) $r$- and (axial) $z$-directions \citep{BAH87}:
\begin{subequations}\begin{align}
          0&=\frac{1}{r}\frac{\partial }{\partial r}(r{\tau_{rr}})+\frac{\partial \tau_{zr}}{\partial z}-\frac{\partial p}{\partial r},
     \label{eq:mom_balance_r}\\
     0&=\frac{1}{r}\frac{\partial }{\partial r}(r{\tau_{rz}})+\frac{\partial \tau_{zz}}{\partial z}-\frac{\partial p}{\partial z}.
     \label{eq:mom_balance_z}     
\end{align}\label{eq:mom_bal_fluid}\end{subequations}
Here, $\boldsymbol{\tau}$ is the fluid's shear stress tensor, and $p$ is the hydrodynamic pressure. We have already made use of the assumptions of axisymmetry in Eqs.~\eqref{eq:mom_bal_fluid}, by neglecting the circumferential stress and any derivatives with respect to $\theta$. The same assumption also leads to the momentum equation in $\theta$ direction being reduced to zero \emph{identically}.

{Now, we need to express the rate-of-strain tensor $\dot{\boldsymbol{\gamma}}$ in terms of the velocity components. For axisymmetric flow with no swirl, the only non-vanishing rate-of-strain components are
\begin{subequations}\begin{alignat}{3}
\dot{\gamma}_{rr} &= \frac{\partial v_r}{\partial r} &&= \frac{\lambda \mathcal{V}_z}{a}\frac{\partial \bar{v}_r}{\partial \bar{r}},\\
\dot{\gamma}_{zz}& = 2\frac{\partial v_z}{\partial z} &&= 2\frac{\lambda \mathcal{V}_z}{a}\frac{\partial \bar{v}_{\bar{z}}}{\partial \bar{z}}, \\
\dot{\gamma}_{rz}&= \dot{\gamma}_{zr} = \frac{\partial v_r}{\partial z} + \frac{\partial v_z}{\partial r}  &&= \frac{\lambda^2 \mathcal{V}_z}{a}\frac{\partial \bar{v}_r}{\partial \bar{z}} + \frac{\mathcal{V}_z}{a} \frac{\partial \bar{v}_{\bar{z}}}{\partial \bar{r}}, \label{eq:dim_gamma_dot_r}
\end{alignat}\label{eq:dim_gamma_dot}\end{subequations}
having used the dimensionless variables from Eq.~\eqref{eq:nd_vars_tube}. Clearly, $(a/\mathcal{V}_z)\dot{\gamma}_{zz} =\mathcal{O}(\lambda)$ and $(a/\mathcal{V}_z)\dot{\gamma}_{rr} =\mathcal{O}(\lambda)$, while $(a/\mathcal{V}_z)\dot{\gamma}_{rz} = (a/\mathcal{V}_z)\dot{\gamma}_{zr}$ have one $\mathcal{O}(1)$ term. Therefore, to the leading order in $\lambda$, the rate-of-strain tensor has two components, consistent with the kinematic reduction to unidirectional flow; $v_r$ and its derivatives, as well as $\partial v_z/\partial z$, are negligible.

Next, we move onto the constitutive equations for the fluid under consideration. Keeping biofluid mechanics applications in mind, we consider the fluid to be non-Newtonian \citep{PKLH99}. Blood is known to exhibit shear-dependent viscosity at steady state, and it is often modeled as a \emph{Casson fluid}, which captures both a yields stress and a shear-dependent viscosity \cite[Ch.~3]{F93}. However, detecting the yield stress (at zero shear rate) in a suspension of blood cells is extremely difficult (perhaps even ``controversial'' \citep{BFO14}), and some experiments \citep{Chien66} show it to be vanishing; see also \citep[p.~65]{F93}. Therefore, we consider the special case of zero yield stress, which reduces the Casson fluid model to the \emph{power-law fluid} (also known as Ostwald--de Waele \citep{Bird76}) model, which connects the stress tensor ${\boldsymbol{\tau}}$ to the rate-of-strain tensor $\dot{\boldsymbol{\gamma}}$ as
\begin{equation}
  {\boldsymbol{\tau}} =  \eta \dot{\boldsymbol{\gamma}}.
\end{equation}
For an incompressible shear flow, the apparent viscosity $\eta$ is a function of the invariants of the rate-of-strain tensor \citep{BAH87}. Specifically, it depends solely on the second invariant $\frac{1}{2}I\!I$ (see, e.g., \citep[\S8.8]{D80}):
\begin{equation}
\label{eq:eta_defined}
    \eta(\dot{\boldsymbol{\gamma}}) = m \left|\frac{1}{2}I\!I\right|^{(n-1)/2},
\end{equation}
where $m$ is the \emph{consistency factor} (a non-negative quantity), and $n$ is the \emph{power-law index} (also a non-negative quantity). Under the condition of axisymmetry, the second invariant takes the form
\begin{equation}\begin{aligned}
    \frac{1}{2}I\!I & = 2 \left(\frac{\partial v_r}{\partial r}\right)^2+\left(\frac{\partial v_z}{\partial z}\right)^2 + \left(\frac{\partial v_r}{\partial z}+\frac{\partial v_z}{\partial r}\right)^2 \\
    & = 2\left(\frac{\lambda \mathcal{V}_z}{a}\right)^2\left[\left(\frac{\partial \bar{v}_{\bar{r}}}{\partial \bar{r}}\right)^2+\left(\frac{\partial \bar{v}_{\bar{z}}}{\partial \bar{z}}\right)^2\right] + \left(\frac{\mathcal{V}_z}{a}\right)^2\left(\lambda^2 \frac{\partial \bar{v}_{\bar{r}}}{\partial \bar{z}} +  \frac{\partial \bar{v}_{\bar{z}}}{\partial \bar{r}}\right)^2 \\
    &= \left(\frac{\mathcal{V}_z}{a}\right)^2 \left(\frac{\partial \bar{v}_{\bar{z}}}{\partial \bar{r}} \right)^2 + \mathcal{O}(\lambda^2).
\end{aligned}\end{equation}

Now, the shear stress components are
\begin{equation}
\label{eq:shear_stress_definition}
    {{\tau_{rz}}=\tau_{zr}} =  \eta(\dot{\boldsymbol{\gamma}}) \left(\frac{\partial v_z}{\partial r} + \frac{\partial v_r}{\partial z}\right).
\end{equation}
Introducing $\mathcal{T}_c$ as the scale for the shear stress, Eq.~\eqref{eq:shear_stress_definition} can be written in dimensionless form:
\begin{equation}
\label{eq:shear_stress_dimless}
\begin{aligned}
  \tau_{rz} = \mathcal{T}_c\bar{\tau}_{\bar{r}\bar{z}} = \tau_{zr} = \mathcal{T}_c \bar{\tau}_{\bar{z}\bar{r}} &= m \left|\frac{1}{2}I\!I\right|^{(n-1)/2}\left(\frac{\lambda^2 \mathcal{V}_z}{a}\frac{\partial \bar{v}_r}{\partial \bar{z}}+\frac{\mathcal{V}_z}{a} \frac{\partial \bar{v}_{\bar{z}}}{\partial \bar{r}}\right)\\
  &= m\left(\frac{\mathcal{V}_z}{a}\right)^n\left|\frac{\partial \bar{v}_{\bar{z}}}{\partial \bar{r}}\right|^{n-1} \frac{\partial \bar{v}_{\bar{z}}}{\partial \bar{r}} + \mathcal{O}(\lambda^{n-1}),
\end{aligned}
\end{equation}
which suggest the choice of stress scale:
\begin{equation}
\label{eq:shear_stress_scale}
    \mathcal{T}_c = m\left(\frac{\mathcal{V}_z}{a}\right)^n.
\end{equation}
Similarly, to the leading order in $\lambda$, the normal stress components are
\begin{subequations}
\begin{align}
  \bar{\tau}_{\bar{z}\bar{z}} &= 
  \lambda \mathcal{T}_c \left|\frac{\partial \bar{v}_{\bar{z}}}{\partial \bar{r}}\right|^{n-1} \frac{\partial \bar{v}_{\bar{z}}}{\partial \bar{z}},\\
  \bar{\tau}_{\bar{r}\bar{r}} &= 
  \lambda \mathcal{T}_c \left|\frac{\partial \bar{v}_{\bar{z}}}{\partial \bar{r}}\right|^{n-1} \frac{\partial \bar{v}_{\bar{r}}}{\partial \bar{r}},
  \end{align}
\end{subequations}
which are of order $\mathcal{O}(\lambda)$ and, clearly, negligible compared to the shear stress components in Eq.~\eqref{eq:shear_stress_dimless}. 

Next, we nondimensionalize the $z$-momentum equation~\eqref{eq:mom_balance_z}:
\begin{equation}
\label{eq:z_momentum_dimless_1}
    0= \mathcal{T}_c \frac{1}{\bar{r}}\frac{\partial }{\partial \bar{r}}(\bar{r}{\bar{\tau}_{\bar{r}\bar{z}}}) + \lambda^2 \mathcal{T}_c\frac{\partial  \bar{\tau}_{\bar{z}\bar{z}}}{\partial \bar{z}} - \lambda\mathcal{P}_c\frac{\partial \bar{p}}{\partial \bar{z}}
\end{equation}
and observe that this equation has a leading-order balance only if \begin{equation}
\label{eq:pressure_scale_defined_shear}
    \mathcal{P}_c =  \frac{\mathcal{T}_c}{\lambda},
\end{equation}
i.e., $\mathcal{P}_c \gg \mathcal{T}_c$, as expected under lubrication theory \cite[Ch.~21]{panton}. It follows that the gradient of the normal stress is negligible to the leading order in $\lambda \ll 1$. 
 
Next, we employ all the information deduced so far to render the radial momentum equation~\eqref{eq:mom_balance_r} dimensionless:
\begin{equation}
\label{eq:mom_balance_r_nondim}
    0=\lambda^2 \mathcal{T}_c\frac{1}{\bar{r}}\frac{\partial }{\partial \bar{r}}(\bar{r}{\bar{\tau}_{\bar{r}\bar{r}}})+\lambda^2 \mathcal{T}_c\frac{\partial  \bar{\tau}_{\bar{r}\bar{z}}}{\partial \bar{z}}-\mathcal{T}_c\frac{\partial \bar{p}}{\partial \bar{r}},
\end{equation}
where we have used Eq.~\eqref{eq:pressure_scale_defined_shear} to replace $\mathcal{P}_c$ with $\mathcal{T}_c/\lambda$. No undetermined scales remain to attempt to balance the last equation; to the leading order in $\lambda$, it simply becomes
\begin{equation}
\label{eq:mom_balance_r_2}
    0=\frac{\partial \bar{p}}{\partial \bar{r}}.
\end{equation}
Therefore, the pressure is  a function of neither $\bar{r}$, from the last equation, nor $\theta$, by assumption~2, hence it is at most a function of $\bar{z}$, i.e., $\bar{p}=\bar{p}(\bar{z})$.  From Eqs.~\eqref{eq:pressure_scale_defined_shear} and \eqref{eq:shear_stress_scale}, we can re-express $\mathcal{P}_c$ in terms of $\mathcal{V}_z$ as
\begin{equation}
\mathcal{P}_c = \frac{m\ell \mathcal{V}_z^n}{a^{n+1}}.
\label{eq:v_p_scale_rel}
\end{equation}
In a flow-rate-controlled experiment/simulation, we can choose a velocity scale $\mathcal{V}_z = q/(\pi a^2)$ based on the constant inlet flow rate $q$, then $\mathcal{P}_c ={m\ell q^n}/{(a^{3n+1}\pi^n)}$.
  
For axisymmetric flow, we expect that the axial velocity will attain its maximum along the centerline ($r=0$), decreasing with the radius until it reaches zero at the tube wall (due to no slip) in this steady flow. Consequently, the velocity gradient is negative and $\left|{\partial \bar{v}_{\bar{z}}}/{\partial \bar{r}}\right|= - {\partial \bar{v}_{\bar{z}}}/{\partial \bar{r}}$. Then, Eq.~\eqref{eq:shear_stress_dimless} becomes
\begin{equation}
\bar{\tau}_{\bar{r}\bar{z}}= -\left(-\frac{\partial \bar{v}_{\bar{z}}}{\partial \bar{r}}\right)^n.
\label{eq:tau_rz}
\end{equation} 
Substituting the latter expression for into the $z$-momentum equation~\eqref{eq:z_momentum_dimless_1}, having neglected the normal stresses of $\mathcal{O}(\lambda^2)$, yields}
\begin{equation}
\frac{1}{\bar{r}}\frac{\partial}{\partial \bar{r}}\left[\bar{r}{\left(-\frac{\partial \bar{v}_z}{\partial \bar{r}}\right)^n}\right] = - \frac{\rd \bar{p}}{\rd \bar{z}}.
\label{eq:non_dim_simplified_momentum}
\end{equation}
 Next, substituting Eq.~\eqref{eq:v_p_scale_rel} into Eq.~\eqref{eq:non_dim_simplified_momentum} and integrating the resulting equation with respect to $\bar{r}$ and requiring that $\bar{v}_{\bar{z}}$ be finite along the centerline, as well as enforcing no slip along the tube's inner wall, $\bar{v}_{\bar{z}} (\bar{r}=\bar{R}) =0$, yields
\begin{equation}
\label{eq:barVz_tube}
 \bar{v}_{\bar{z}} = \left(-\frac{1}{2}\frac{\rd \bar{p}}{\rd \bar{z}}\right)^{1/n}\left(\frac{\bar{R}^{1+1/n}-\bar{r}^{1+1/n}}{1+1/n}\right),
\end{equation}
where $\bar{R} = R/a$ is the dimensionless deformed tube radius. Note that $\bar{R}$ is not necessarily unity because we allow the tube to deform due to FSI, as discussed in the next section. As a result, while $\bar{p}$ is at most a function of $\bar{z}$, $\bar{v}_{\bar{z}}$ can depend upon both $\bar{r}$ and $\bar{z}$.

\section{Structural mechanics problem}
\label{section:deformation}

In \S\ref{sec:flow}, the momentum conservation equations for a power-law fluid were reduced to unidirectional flow. They explicitly depend only on the radial coordinate, up to the leading order under the assumptions of axisymmetry and the lubrication approximation. In a similar manner, we now model the structural mechanics of the tube. To make the problem tractable analytically, the equations stating the equilibrium of forces in the solid are simplified under the following assumptions:
\begin{enumerate}
  \item \label{point:1}The tube is thin; its thickness is negligible compared to its radius: $t\ll a$.
  \item \label{point:2}The tube is slender; its radius is small compared to its length: $a\ll \ell$.
  \item \label{point:3} The material from which the tube is composed is isotropic and linearly elastic, with elasticity (Young's) modulus $E$ and Poisson ratio $\nu$, so the relationship between stress and strain is linear. 
  \item \label{point:4} The strains are small, so the relationship between strain and displacement is linear.
  \item \label{point:5} The characteristic radial deformation $\mathcal{U}_c$ is small compared to the (smallest) characteristic dimension of the tube: $\mathcal{U}_c \ll t$.
\end{enumerate}
Here, assumptions \ref{point:3} and \ref{point:4} ensure that the relation between stress and displacement is linear. Thus, the corresponding theory developed in this paper pertains to what we shall term \emph{linear} FSI. The ramifications of assumptions \ref{point:1} and \ref{point:2} will be discussed in the context of shell theories. Assumption~\ref{point:5} implies our theory is a \emph{small-deformation} FSI theory, thus we may work the problem in Eulerian coordinates. {In general, soft elastic structures (e.g., blood vessels) can exhibit a nonlinear material (hyperelastic) response, as well as viscoelasticity and even anisotropy (see, e.g., \cite[Ch.~8 and 9]{F93}). Therefore, the linear FSI theory developed in the present work, under the above five assumptions, must be understood as the simplest mathematical model with the key FSI features.}

A shell theory models the dynamics of a 3D entity in two dimensions, thus it by definition an approximate theory. Approximations are introduced in every facet of shell theory: in the strain--displacement relation (\emph{kinematics}), in the stress equilibrium relation (\emph{statics}), and in the stress--strain relation (\emph{constitutive}). There are many shell theories, of varying degree of approximation, as discussed in the classic monographs by Kraus~\citep{Kraus67}, Flugge~\citep{Flugge60}, and Timoshenko and Woinowsky-Krieger~\citep{TWK59}. We focus only on the ``simplest'' shell theories capable of describing the FSI problem posed above.

\subsection{Membrane theory}
\label{sec:membrane}

The thinness assumption ($t\ll a$) allows us to analyze  the tube using membrane theory for sufficiently small $t/a$. Membrane theory of shells pertains to structures that sustain only tension (in the axial and/or in the circumferential directions) but cannot support bending or twisting moments \cite[Ch.~3]{Flugge60}. Furthermore, the radial stress developed inside the structure is negligible, at the leading order in $t/a$, compared to the hoop and axial stress.

Owing to the slenderness of the geometry, i.e., $\lambda = a/\ell \ll 1$, we start our exposition of the membrane theory by assuming a state of plane strain. In other words, $\varepsilon_{zz} = \partial u_z/\partial z =0$. Next, since the edges are clamped, we conclude that $u_z =0$ along the length of the tube. Indeed, most tubes in physiology, like arteries, tracheoles, urethra are longitudinally constrained \textit{in situ} \citep{VON11,F97}. Note that neglecting the axial displacement is a fairly common assumption in the hemodynamics literature \citep{NV99,QTV00,FGNQ01,CLMT05}; for a more rigorous derivation, see \cite{CGM07}.

From the constitutive equation of linear elasticity [see, e.g., Eqs.~(3.17) and (3.18a) in \cite{Flugge60}], we then have:
\begin{equation}
\label{eq:equal_stress_uz_zero}
  \varepsilon_{zz} = \frac{1}{E}(\sigma_{zz}-\nu\sigma_{\theta\theta}) = 0 \qquad \Rightarrow \qquad \sigma_{zz} = \nu\sigma_{\theta\theta},
\end{equation}
where $\sigma_{\theta\theta}$ is the hoop stress, while $\sigma_{zz}$ and $\varepsilon_{zz}$ are the axial stress and strain, respectively. To find an expression for $\sigma_{\theta\theta}$, we appeal to the equation of static equilibrium in the radial direction:
\begin{equation}
N_{\theta} = a p,
\label{eq:Ntheta_p}
\end{equation}
where $N_\theta$ is the stress resultant in the azimuthal direction of the tube. Here, the hydrodynamic pressure $p(z)$ provides the load, and we have shown in \S\ref{sec:flow} that $p$ is, at most, a function of $z$. Next, membrane theory assumes that due to the cylinder being thin, the stress across the thickness is uniform. Therefore,  $\sigma_{\theta\theta}$ is simply the corresponding stress resultant divided by the cylinder's thickness $t$:
\begin{equation}
\sigma_{\theta\theta} = \frac{N_\theta}{t} = \left(\frac{a}{t}\right)p.
\label{eq:hoop_stress}
\end{equation}

Finally, under the assumption of axisymmetric deformation \citep[Eq.~(3.18b)]{Flugge60} and Eqs.~\eqref{eq:equal_stress_uz_zero} and \eqref{eq:hoop_stress}, we obtain
\begin{equation}
\label{eq:presrure_vessel_radial_main}
 u_r(z) = \varepsilon_{\theta\theta}a = \frac{1}{E}(\sigma_{\theta\theta}-\nu\sigma_{zz})a = (1-\nu^2)\left(\frac{a^2}{Et}\right)p(z)
\end{equation}
for the radial displacement. The most important conclusion to be drawn from the analysis in this subsection is that, under the membrane theory, the tube's radial displacement is simply proportional to the local hydrodynamic pressure, with the geometric and elasticity parameters setting the proportionality constant.

\subsection{Donnell shell theory}
\label{sec:donnell}

From the several theories for thin shells of revolution based on the Love--Kirchhoff hypothesis \citep{L88}, perhaps the earliest and most ``popular'' is  Donnell's shell theory \citep{DonnellShell}. Donnell's shell theory is a straightforward extension of thin-plate theory to shells \citep{Kraus67}, which itself is an extension of  Euler-beam theory to two dimensions. Furthermore, for the special case of axisymmetric loads with zero curvature, Donnell's shell theory reduces identically to the Kirchhoff--Love thin-plate theory \citep{Kraus67}, which we have successfully employed to analyze microchannel FSIs \citep{CCSS17,ADC18}. Improving upon the membrane theory of \S\ref{sec:membrane}, Donnell's shell theory takes into account bending moments and the variation of the stresses across the shell's thickness \citep{Flugge60,Kraus67}.

To be consistent with the membrane theory of \S\ref{sec:membrane}, we again neglect the axial displacement ($u_z\equiv0$). Then, following Dym~\citep[Ch.~V]{D90}, the equation expressing the momentum balance (for axisymmetric deformation and loading) of a Donnell shell is
\begin{equation}
\label{eq:Bending_Donnell_OriginalCoordinates}
\frac{\rd^2 M_z}{\rd z^2} -\frac{N_{\theta}}{a}=-p(z).
\end{equation}
Here, the bending moment $M_z$ is expressed through the linear elastic law as
\begin{equation}
 \label{eq:Bending_Moment_Donnell}
M_z =-K\frac{\rd^2 u_r}{\rd z^2},
\end{equation}
where $K=Et^3/[12(1-\nu^2)]$ is the \emph{bending (flexural) rigidity} of the shell. The stress resultant in the circumferential direction is
\begin{equation}
 \label{eq:Circumferential_Stress_Donnell}
 N_{\theta} = D\left(\frac{u_r}{a}\right),
\end{equation}
where $D=Et/(1-\nu^2)$ is the \emph{extensional rigidity} of the shell.
Then, Eq.~\eqref{eq:Bending_Donnell_OriginalCoordinates}, when written in terms of the displacement using Eqs.~\eqref{eq:Bending_Moment_Donnell} and \eqref{eq:Circumferential_Stress_Donnell}, and simplified by substituting the expressions for $D$ and $K$, becomes an ordinary differential equation (ODE) for the radial deflection $u_r(z)$ forced by the hydrodynamic pressure $p(z)$:
\begin{equation}
  \label{eq:Cylinder_Structure}
  \frac{Et^3}{12(1-\nu^2)} \left( \frac{\rd^4 u_r}{\rd z^4} + \frac{12}{a^2t^2} u_r \right) = p.
\end{equation}

To understand the dominant balance(s) in Eq.~\eqref{eq:Cylinder_Structure}, we introduce dimensionless variables, some of which are restated from Eq.~\eqref{eq:nd_vars_tube}, as follows:
\begin{equation}
  \bar{z} = {z}/{\ell},\qquad \bar{u}_{\bar{r}} = {u_r}/{\mathcal{U}_c}, \qquad \bar{p} = {p}/{\mathcal{P}_c}.
\label{eq:nd_vars_tube2}
\end{equation}
The characteristic scale for the radial deflection of the tube, $\mathcal{U}_c$, is to be determined self-consistently as part of this analysis. Substituting the dimensionless variables from Eq.~\eqref{eq:nd_vars_tube2} into Eq.~\eqref{eq:Cylinder_Structure} yields
\begin{equation}
 \label{eq:structure_deflection_non_dim}
 \left(\frac{t}{a}\right)^2\left(\frac{a}{\ell}\right)^4\frac{\rd^4 \bar{u}_{\bar{r}}}{\rd{\bar{z}}^4} + 12\bar{u}_{\bar{r}} = \frac{12(1-\nu^2) a^2\mathcal{P}_c}{Et\,\mathcal{U}_c}\bar{p}.
\end{equation}
For a thin and slender shell we can neglect, in an order of magnitude sense to the leading order in $t/a$ and $a/\ell$, the first term on the left-hand side of Eq.~\eqref{eq:structure_deflection_non_dim} to obtain:
\begin{equation}
 \bar{u}_{\bar{r}} = \frac{a^2\mathcal{P}_c}{E t \,\mathcal{U}_c}(1-\nu^2)\bar{p}.
\label{mt_structure_non_dim}
\end{equation}
Since Eq.~\eqref{mt_structure_non_dim} represents a leading-order balance, we are free to choose the deformation scale in terms of the pressure scale as 
\begin{equation}
 \mathcal{U}_c = \frac{a^2\mathcal{P}_c}{Et}.
\label{mt_deflection_scale}
\end{equation}
Hence, the deformed tube radius is
\begin{equation}
 \bar{R}(\bar{z}) \equiv \frac{a+u_r(z)}{a} = 1 + \beta\bar{u}_{\bar{r}}(\bar{z}),
\label{eq:R1_nd}
\end{equation}
where $\beta:=\mathcal{U}_c/a$ is a dimensionless parameter that ``controls'' the fluid--structure interaction. It is a measure of the magnitude of the characteristic radial deformation $\mathcal{U}_c$ compared to the undeformed radius $a$. A larger value of $\beta$ corresponds to ``stronger'' fluid--structure coupling and, thus, a larger deformation.

Thus, at the leading order in $t/a$ and $a/\ell$, Eq.~\eqref{eq:structure_deflection_non_dim} yields a simple  deformation--pressure relation:
\begin{equation}
 \label{nondimensional_reduced_deflection_microtube}
 \bar{u}_{\bar{r}}(\bar{z}) = (1-\nu^2)\bar{p}(\bar{z}).
\end{equation}
Note that Eq.~\eqref{nondimensional_reduced_deflection_microtube} is identically the dimensionless version of our membrane theory result in  Eq.~\eqref{eq:presrure_vessel_radial_main}. Also, in obtaining Eq.~\eqref{nondimensional_reduced_deflection_microtube}, we have \emph{singularly perturbed} Eq.~\eqref{eq:structure_deflection_non_dim}, a point that we revisit in \S\ref{sec:beyond_leading_order_fsi}.

\begin{remark}
Mathematical analogues to  Eq.~\eqref{eq:Cylinder_Structure} exist in at least two different domains of structural mechanics. First is the governing equation of bending of a beam placed on an elastic foundation due to Winkler~\citep{W67} (see also the book \citep{H46} and recent review article  \citep{DMKBF18}). The force generated by the elastic foundation is directly proportional to the local displacement of the beam and this leads to the linear term in Eq.~\eqref{eq:Cylinder_Structure}. The fourth-order derivative term is due to the beam's bending resistance. Second, is the governing equation of the quasi-static  planar spreading of a fluid under an elastic beam (ignoring tension) driven by gravity---a problem that arises in geophysical fluid dynamics; see \citep[Eq.~(4.2)]{HBDB14} and \citep{BC11,BN18}. In this example, the fourth-order derivative term is again due to bending resistance, but the term that is linear in the displacement arises from the hydrostatic pressure due to gravity.
\end{remark}

\begin{remark}
To the leading order in $t/a$ and  $a/\ell$, bending in Donnell's shell theory is negligible and this theory leads to the same result as the membrane theory, namely Eq.~\eqref{eq:presrure_vessel_radial_main} [and its dimensionless counterpart, Eq.~\eqref{nondimensional_reduced_deflection_microtube}], which dictates that the radial deflection of the tube is directly proportional to the pressure at a given flow-wise cross-section. This result is also in agreement with the results of Elbaz and Gat~\citep{EG14,EG16}, taking into account, of course, the different boundary conditions employed therein.
\end{remark}

\section{Coupling the fluid mechanics and structural problems: Flow rate--pressure drop relation}
\label{section:coupling}
We now turn to the main task, which is evaluating the pressure drop and thus generalizing the Hagen--Poiseuille law to deformable tubes. The flow rate in the tube is by definition
\begin{equation}
q = \int\limits_{0}^{2\pi}\int\limits_{0}^{R(z)}v_z(r,z) \, r \,\rd r\, \rd\theta 
= \mathcal{V}_z 2\pi a^2 \int\limits_{0}^{{\bar{R}(\bar{z})}}{\bar{v}}_{\bar{z}}(\bar{r},\bar{z}) \, \bar{r} \,\rd\bar{r},
\label{eq:q_def}
\end{equation}
where the second equality follows from performing the (trivial) azimuthal integration and introducing the dimensionless variables from Eq.~\eqref{eq:nd_vars_tube}.  Now, substituting the expression for 
$\bar{v}_{\bar{z}}$ from Eq.~\eqref{eq:barVz_tube} into Eq.~\eqref{eq:q_def} yields:
\begin{equation}
\frac{q}{\mathcal{V}_z \pi a^2} =  \left(-\frac{1}{2}\frac{\rd\bar{p}}{\rd\bar{z}}\right)^{1/n}\frac{[\bar{R}(\bar{z})]^{3+1/n}}{3+1/n}.
\end{equation}
In a steady incompressible flow, conservation of mass requires that the volumetric flow rate is a constant independent of $z$. Then, owing to our choice of axial velocity scale, $q/(\mathcal{V}_z\pi a^2)=1$ [recall the discussion after Eq.~\eqref{eq:v_p_scale_rel}], and the above equation can be rewritten as:
\begin{equation}
\frac{\rd \bar{p}}{\rd \bar{z}} = -2[(3+1/n)]^n [\bar{R}(\bar{z})]^{-(3n+1)}.
\label{recipe_for_ODE_numerical}
\end{equation}
This is an ODE for $\bar{p}(\bar{z})$, subject to an appropriate closure relation for $\bar{R}(\bar{z})$.

\subsection{Rigid tube}

First, for completeness and future reference, consider the case of $\bar{R}=1$ (rigid tube of uniform radius). Equation \eqref{recipe_for_ODE_numerical} can be immediately integrated, subject to an outlet boundary condition [$\bar{p}(1)=0$], to yield the usual linear pressure profile:
\begin{equation}
\bar{p} (\bar{z}) = 2[(3+1/n)]^n(1-\bar{z}).
\label{eq:pz_no_fsi}
\end{equation}
Since it is our convention that $\bar{p}(\bar{z}=0)=\Delta \bar{p}$ is the full pressure drop across the length of the tube, then $\Delta \bar{p} = 2[(3+1/n)]^n$, which is the well-known {Hagen--Poiseuille law for a power-law fluid} \citep{BAH87}.

\subsection{Leading-order-in-thickness (membrane) theory}
\label{sec:leading_order_fsi}

Next, inserting the relation $\bar{R} = 1 + (1-\nu^2)\beta\bar{p}$ [having employed Eqs.~\eqref{eq:R1_nd} and \eqref{nondimensional_reduced_deflection_microtube}] into Eq.~\eqref{recipe_for_ODE_numerical} yields:
\begin{equation}
\frac{\rd \bar{p}}{\rd \bar{z}} = -2[(3+1/n)]^n[1+(1-\nu^2)\beta \bar{p}]^{-(3n+1)}.
\label{eq:dpdz_leading_order}
\end{equation}
Separating variables and integrating subject to $\bar{p}(1)=0$, we have:
\begin{equation}
\bar{p}(\bar{z}) = \frac{1}{(1-\nu^2)\beta}\left\{\left[1 + 2(3n+2)(1-\nu^2)\beta[(3+1/n)]^n (1-\bar{z})\right]^{1/(3n+2)} - 1\right\}.
\label{eq:P_vs_Z_Microtube}
\end{equation}
Again, the full dimensionless pressure drop is obtained by evaluating Eq.~\eqref{eq:P_vs_Z_Microtube} at $\bar{z}=0$:
\begin{equation}
\label{eq:DP_vs_Q_Microtube}
 \Delta\bar{p}=\frac{1}{(1-\nu^2)\beta}\left\{\left[1+2(3+1/n)^n(3n+2)(1-\nu^2)\beta\right]^{1/(3n+2)}-1\right\} .
\end{equation} 
Notice that Eqs.~\eqref{eq:P_vs_Z_Microtube} and \eqref{eq:DP_vs_Q_Microtube} are explicit relations for $\bar{p}$ and $\Delta\bar{p}$, respectively, which is unlike the case of microchannels \citep{CCSS17,SC18,ADC18}.

\begin{remark}
Equation~\eqref{eq:dpdz_leading_order}, when written in dimensional form, can be inverted to yield the flow rate in terms of the pressure gradient: 
\begin{equation}
{q} = \varsigma({p}) \left(-\frac{\rd {p}}{\rd {z}}\right)^{1/n},\qquad \varsigma({p}) := \frac{1}{3+1/n}\left( \frac{a}{2m}\right)^{1/n}\left[1+(1-\nu^2) \frac{a}{Et}p\right]^{3+1/n}.
\end{equation}
This equation is, clearly, a generalization of the classic result of Rubinow and Keller~\citep{RK72} for steady low $Re$, Newtonian flow in a deformable tube. Importantly,  we have self-consistently derived the function $\varsigma(p)$ that accounts for steady non-Newtonian FSI in a tube.
\end{remark}

\begin{remark} The maximum radial displacement of the tube wall over its length is $\beta\bar{u}_{\bar{r}}(0)$ [recall Eq.~\eqref{eq:R1_nd}]. Using Eqs.~\eqref{nondimensional_reduced_deflection_microtube} and \eqref{eq:P_vs_Z_Microtube}, it can then  computed to be
\begin{equation}
\max_{0\le\bar{z}\le1} \beta\bar{u}_{\bar{r}}(\bar{z}) = \big\{1 + 2(3n+2)(1-\nu^2)\beta [(3+1/n)]^n \big\}^{1/(3n+2)} - 1.
\end{equation}
Note that, once the solid and fluid properties ($\nu$ and $n$) are fixed, the maximum displacement is solely a function of the FSI parameter $\beta$.
\end{remark}

\begin{remark} To perform the consistency check of recovering the rigid-tube pressure profile in Eq.~\eqref{eq:pz_no_fsi} as the $\beta\to0^+$ limit of the deformable-tube pressure profile in Eq.~\eqref{eq:P_vs_Z_Microtube}, we must realize that $\beta\to0^+$ in Eq.~\eqref{eq:P_vs_Z_Microtube} is a ``$0/0$'' limit. L'H\^{o}pital's rule or a Taylor series in $\beta\ll1$ easily shows that 
Eq.~\eqref{eq:pz_no_fsi} is indeed the $\beta\to0^+$ limit of Eq.~\eqref{eq:P_vs_Z_Microtube}.
\end{remark}

\begin{remark}
For the special case of a Newtonian fluid (i.e., $n=1$ and $m=\mu$), Eq.~\eqref{eq:P_vs_Z_Microtube} reduces to:
\begin{equation}
\bar{p}(\bar{z}) = \frac{1}{(1-\nu^2)\beta} \left\{\left[ 1 + 40(1-\nu^2)\beta (1-\bar{z})\right]^{1/5}-1\right\}.
\label{eq:P_vs_Z_Microtube_Newtonian}
\end{equation}
A Taylor series expansion in $\beta\ll1$ of Eq.~\eqref{eq:P_vs_Z_Microtube_Newtonian} yields the pressure distribution in Poiseuille flow: $\bar{p}(\bar{z}) = 8(1-\bar{z}) + \mathcal{O}(\beta)$.
\end{remark}


\subsection{Beyond leading-order-in-thickness theory}

\label{sec:beyond_leading_order_fsi}

In \S\ref{sec:leading_order_fsi}, we obtained the flow rate--pressure drop relationship considering only the leading-order deformation profile as given by Eq.~\eqref{nondimensional_reduced_deflection_microtube}. In this subsection, we venture beyond the leading-order approximation by solving the ``full'' ODE, namely Eq.~\eqref{eq:structure_deflection_non_dim} for the deformation under Donnell's shell theory. Equation~\eqref{eq:structure_deflection_non_dim} is coupled to Eq.~\eqref{recipe_for_ODE_numerical}, which relates the pressure gradient and the tube's radial deformation. Taking $\rd/\rd \bar{z}$ of Eq.~\eqref{eq:structure_deflection_non_dim}, eliminating $\rd \bar{p}/\rd \bar{z}$ using Eq.~\eqref{recipe_for_ODE_numerical}, substituting $\bar{R}$ and $\mathcal{U}_c$ from Eqs.~\eqref{eq:R1_nd} and \eqref{mt_deflection_scale}, respectively, we obtain a single \emph{nonlinear fifth-order} ODE in the deformation:
\begin{equation}
\left(1+\beta\bar{u}_{\bar{r}}\right)^{3n+1} \Bigg[ \underbrace{\left(\frac{t}{a}\right)^2\left(\frac{a}{\ell}\right)^4\frac{\rd^5\bar{u}_{\bar{r}}}{\rd{\bar{z}}^5}}_{\text{bending}} + \underbrace{12\frac{\rd\bar{u}_{\bar{r}}}{\rd{\bar{z}}}}_{\text{stretching}} \Bigg]
= \underbrace{-24(1-\nu^2)^2 [{(3+1/n)}]^n}_{\text{loading}}.
\label{eq:coupled_ODE_non_dim}
\end{equation}
The ODE~\eqref{eq:coupled_ODE_non_dim} is subject to the following boundary conditions expressing clamping of the shell at the inlet and outlet planes [Eqs.~\eqref{eq:coupledBC_z0} and \eqref{eq:coupledBC_z1}, respectively] and zero gauge pressure at the outlet [Eq.~\eqref{eq:coupledBC_p}]:
\begin{subequations}\begin{align}
  \bar{u}_{\bar{r}}|_{\bar{z}= 0} = \left.\frac{\rd \bar{u}_{\bar{r}}}{\rd \bar{z}}\right|_{\bar{z}= 0} &= 0, \label{eq:coupledBC_z0}\displaybreak[3]\\
  \bar{u}_{\bar{r}}|_{\bar{z}=1} = \left. \frac{\rd \bar{u}_{\bar{r}}}{\rd \bar{z}}\right|_{\bar{z}=1} &= 0, \label{eq:coupledBC_z1}\displaybreak[3]\\
  \left. \frac{\rd^4 \bar{u}_{\bar{r}}}{\rd \bar{z}^4}\right|_{\bar{z}=1} &= 0. \label{eq:coupledBC_p}
\end{align}\label{eq:coupledBC}\end{subequations}
Equations~\eqref{eq:coupled_ODE_non_dim} and \eqref{eq:coupledBC} represent a nonlinear \emph{two-point} (TP) \emph{boundary value problem} (BVP) \citep{Keller76}, the solution of which fully characterizes the physics of steady FSI in an elastic tube conveying a non-Newtonian fluid.

As alluded to in \S\ref{sec:donnell}, for a slender ($a \ll \ell$) and thin ($t \ll a$) structure, Eq.~\eqref{eq:coupled_ODE_non_dim} becomes an example of a \textit{singular perturbation problem} in the limit of vanishing $t/a$ and $a/\ell$. Physically, ``boundary layers'' develop near the inlet and outlet of the tube, where the bending due to clamping becomes significant, as opposed to the rest of the tube where stretching dominates. A similar story unfolds for gravity-driven spreading of a viscous fluid under an elastic beam \citep{BN18}: an elasto-gravity length scale divides the domain into an inner region, in which pressure is hydrostatic and due to gravity, and a peripheral region, in which bending is also important.

As a singular perturbation problem, Eq.~\eqref{eq:coupled_ODE_non_dim} is now amenable to treatment via \emph{matched asymptotics} \cite[Ch.~2]{H13}. First, we introduce a dimensionless small parameter $\epsilon = \sqrt{{ta}/{\ell^2}} \ll 1$, then Eq.~\eqref{eq:coupled_ODE_non_dim} can be rewritten as:%
\begin{equation}
\label{eq:coupled_ODE_non_dim_epsilon}
\left(1+\beta\bar{u}_{\bar{r}}\right)^{3n+1} \left(\epsilon^4\frac{\rd^5\bar{u}_{\bar{r}}}{\rd{\bar{z}}^5}+12\frac{\rd\bar{u}_{\bar{r}}}{\rd{\bar{z}}}\right) = -24(1-\nu^2)^2 [{(3+1/n)}]^n.
\end{equation}
As is standard, we first let $\epsilon\to0^+$, thus singularly perturbing the ODE, and obtain the governing equation for the solution in the \emph{outer region}:
\begin{equation}
\label{eq:coupled_ODE_non_dim_outer}
\left(1 + \beta\bar{u}_{\bar{r}}\right)^{3n+1} \frac{\rd\bar{u}_{\bar{r}}}{\rd{\bar{z}}} = -2(1-\nu^2)^2 [{(3+1/n)}]^n.
\end{equation}
The outer solution must ``respect'' the first part of the clamping condition at $\bar{z}=1$, i.e., $\bar{u}_{\bar{r}}|_{\bar{z}=1}=0$ from Eq.~\eqref{eq:coupledBC_z1}. Then, the solution to the ODE~\eqref{eq:coupled_ODE_non_dim_outer} is
\begin{equation}
\bar{u}_{\bar{r}} (\bar{z}) = \frac{1}{\beta}\left(\left\{ 1 + 2\beta (3n+2)(1-\nu^2)^2 [{(3+1/n)}]^n (1-\bar{z}) \right\}^{1/(3n+2)} - 1 \right).
\label{eq:outer_ode_sol}
\end{equation}
Note that Eq.~\eqref{eq:outer_ode_sol} can also be obtained by combining Eqs.~\eqref{nondimensional_reduced_deflection_microtube} and \eqref{eq:P_vs_Z_Microtube} from the membrane theory, showing the consistency of our two structural mechanics models.

To satisfy the clamped boundary condition at $\bar{z}=0$, i.e., Eq.~\eqref{eq:coupledBC_z0}, we must introduce a \emph{boundary layer} near $\bar{z}=0$ wherein the highest-order derivative in Eq.~\eqref{eq:coupled_ODE_non_dim_epsilon} is dominant and is retained. Then, an inner solution can be obtained. To this end, we introduce a scaled spatial coordinate $\zeta$ such that for $\bar{z}\ll1$, $\zeta = \mathcal{O}(1)$. A straightforward balancing argument leads us to define $\zeta = \bar{z}/\epsilon$. Then, the nonlinear ODE~\eqref{eq:coupled_ODE_non_dim_epsilon} becomes
\begin{equation}
\label{eq:coupled_ODE_non_dim_epsilon_inner}
\frac{\rd^5\bar{u}_{\bar{r}}}{\rd{{\zeta}}^5}+{12}\frac{\rd\bar{u}_{\bar{r}}}{\rd{\zeta}} = -\epsilon \left\{\frac{24(1-\nu^2)^2 [{(3+1/n)}]^n}{\left(1 + \beta\bar{u}_{\bar{r}}\right)^{3n+1} }\right\}.
\end{equation}
At the leading order in $\epsilon\ll 1$, we have
\begin{equation}
\label{eq:ODE_Boundary_layer}
\frac{\rd^5\bar{u}_{\bar{r}}}{\rd{{\zeta}}^5}+{12}\frac{\rd\bar{u}_{\bar{r}}}{\rd{\zeta}} = 0
\end{equation}
subject to the following boundary conditions:
\begin{subequations}\begin{align}
  \bar{u}_{\bar{r}}|_{\zeta= 0} = 
  \left.\frac{\rd\bar{u}_{\bar{r}}}{\rd\zeta}\right|_{\zeta= 0} &= 0, \\
  \bar{u}_{\bar{r}}|_{\zeta \rightarrow \infty} &= \left.\bar{u}_{\bar{r}}^\text{outer}\right|_{\bar{z}=0},\label{eq:matchedBC}
\end{align}\end{subequations}
where $\bar{u}_{\bar{r}}^\text{outer}$ denotes the outer solution from Eq.~\eqref{eq:outer_ode_sol}. Here, the first two boundary conditions (at $\zeta=0$) are due to clamping, while the remaining boundary condition (as $\zeta \rightarrow \infty$) is necessary to \textit{match} the inner solution to the outer solution. 

The general solution to Eq.~\eqref{eq:ODE_Boundary_layer} that decays as $\zeta \to \infty$ is
\begin{equation}
\label{eq:ODE_Boundary_layer_integrated5}
\bar{u}_{\bar{r}}(\zeta) = \re^{-\zeta\sqrt[4]{3}} \left[ \tilde{C}_2\sin\left(\zeta\sqrt[4]{3}\right) + \tilde{C}_4\cos\left(\zeta\sqrt[4]{3}\right) \right] + C_0,
\end{equation}
Now, we apply the boundary condition  $\bar{u}_{\bar{r}}|_{\zeta=0}=0$ to obtain $\tilde{C}_4 = - C_0$. Next, we use the boundary condition  $({\rd \bar{u}_{\bar{r}}}/{\rd \zeta})|_{\zeta=0}=0$ to find that $\tilde{C}_2=\tilde{C}_4$. Finally, from the matching condition in Eq.~\eqref{eq:matchedBC}, we find
\begin{equation}
\label{eq:C_4}
\tilde{C}_4 = \frac{1}{\beta} \left( 1-\Big\{ 1+ 2\beta(3n+2)(1-\nu^2)^2 [{(3+1/n)}]^n\Big\}^{1/({3n+2})} \right).
\end{equation}
Thus, the final expression for the inner solution in the boundary layer near $\bar{z}=0$, to the leading order in $\epsilon$, is
\begin{equation}
\label{eq:ODE_Boundary_layer_integrated_final}
\bar{u}_{\bar{r}}(\zeta) \sim \tilde{C}_4 \left\{ \re^{-\zeta\sqrt[4]{3}} \left[ \sin\left(\zeta\sqrt[4]{3}\right) + \cos\left(\zeta\sqrt[4]{3}\right) \right] -1 \right\}.
\end{equation}

Next, another boundary layer must exist near the outlet at $\bar{z}=1$ because, although $\bar{u}_{\bar{r}}^\mathrm{outer}\to0$ as $\bar{z}\to1$, $\rd\bar{u}_{\bar{r}}^\mathrm{outer}/\rd\bar{z}\not\to0$ as $\bar{z}\to1$, i.e., the clamping boundary condition is not fully satisfied. Thus, we expect both the dependent (deformation) and the independent (axial position) variables to be small in this layer. That is, we conjecture that the boundary layer at $\bar{z}=1$ is actually a \emph{corner layer} \citep[\S2.6]{H13} (sometimes termed a \emph{derivative layer} \cite[see][pp.~85--93]{N15}). Indeed, \cite{HBDB14} also observed boundary and corner layers in the related problem of gravity-driven spreading of a viscous fluid under an elastic beam. The boundary layers in their study were the result of the need for ``regularization'' of the contact line at the advancing fluid front, accomplished by introducing a pre-wetting film. Unlike the present model, however, the differential equation in the inner and outer regions, along with the pertinent matching conditions, had to be solved numerically by \cite{HBDB14}, due to their complexity.

Now, introducing the rescalings ${\zeta}=(1-\bar{z})/\epsilon^{\alpha_1}$ and $\hat{u}(\zeta) = \bar{u}_{\bar{r}}(\bar{z})/\epsilon^{\alpha_2}$ into Eq.~\eqref{eq:coupled_ODE_non_dim_epsilon}, we can balance all three terms if and only if $\alpha_1=\alpha_2=1$. The first and last terms can be balanced for any $4-5\alpha_1+\alpha_2=0$ as long as $\alpha_1 < \alpha_2$ but then there is a non-uniqueness of the boundary layer thickness, so we discard this possibility. Thus, the nonlinear ODE~\eqref{eq:coupled_ODE_non_dim_epsilon} becomes
\begin{equation}
\label{eq:coupled_ODE_non_dim_epsilon_corner}
\frac{\rd^5\hat{u}}{\rd\zeta^5}+{12}\frac{\rd\hat{u}}{\rd\zeta} = \frac{24(1-\nu^2)^2 [{(3+1/n)}]^n}{\left(1 + \beta\epsilon\hat{u}\right)^{3n+1}}.
\end{equation}
Expanding in $\epsilon\ll1$, we have, at the leading order,
\begin{equation}
\label{eq:coupled_ODE_non_dim_leading_corner}
\frac{\rd^5\hat{u}}{\rd{\zeta}^5} + 12\frac{\rd\hat{u}}{\rd{\zeta}} = 12\mathcal{A},
\end{equation}
where for convenience we have defined
$\mathcal{A}:=2(1-\nu^2)^2 [{(3+1/n)}]^n$. The ODE~\eqref{eq:coupled_ODE_non_dim_leading_corner} must satisfy the remaining boundary conditions at $\bar{z}=1$, from Eqs.~\eqref{eq:coupledBC_z1} and \eqref{eq:coupledBC_p}, that are not satisfied by the outer solution, namely
\begin{subequations}\begin{align}
  \left.\frac{\rd\hat{u}}{\rd\zeta}\right|_{{\zeta}= 0} = \left.\frac{\rd^4\hat{u}}{\rd\zeta^4}\right|_{{\zeta}= 0} &= 0,\label{eq:coupledBC2} \\
  \hat{u}|_{\zeta \rightarrow \infty} &= \left.\bar{u}_{\bar{r}}^\text{outer}\right|_{\bar{z}=1}.\label{eq:matchedBC2}
\end{align}\end{subequations}

The general solution to Eq.~\eqref{eq:coupled_ODE_non_dim_leading_corner} that decays as $\zeta \to \infty$ is
\begin{equation}
\hat{u}(\zeta) = \re^{-\zeta\sqrt[4]{3}} \left[ \tilde{C}_2\sin\left(\zeta\sqrt[4]{3}\right) + \tilde{C}_4\cos\left(\zeta\sqrt[4]{3}\right) \right] + C_0 + \mathcal{A}\zeta.
\end{equation}
Now, we impose the boundary condition $(\rd \hat{u}/\rd \zeta)|_{\zeta = 0}=0$ to find that
$\tilde{C}_2 = \tilde{C}_4 - \mathcal{A}/\sqrt[4]{3}$.
Finally, the boundary condition $(\rd^4\hat{u}/\rd \zeta^4)|_{\zeta=0}=0$ requires that $\tilde{C}_4 = 0$.
Thus, we have obtained a fully-specified corner layer (inner) solution:
\begin{equation}
\hat{u}(\zeta) \sim \mathcal{A} \left[ \zeta - \frac{\re^{-\zeta\sqrt[4]{3}}}{\sqrt[4]{3}} \sin\left(\zeta\sqrt[4]{3}\right) \right] + C_0.
\label{eq:ODE_corner_layer_integrated_final}
\end{equation}
The inner solution in Eq.~\eqref{eq:ODE_corner_layer_integrated_final} must still be matched to the outer solution in Eq.~\eqref{eq:outer_ode_sol}, which goes to zero as $\bar{z}\to 1$. Thus, we immediately conclude that $C_0 = 0$, and the common part of the inner and outer solutions is $\mathcal{A}\zeta$ [as can be confirmed by a Taylor series expansion of Eq.~\eqref{eq:outer_ode_sol} for $\bar{z}\approx 1$]. This argument can be made even more rigorous using an intermediate variable matching procedure as in \citep[\S2.6]{H13}.

Finally, adding together Eqs.~\eqref{eq:outer_ode_sol}, \eqref{eq:ODE_Boundary_layer_integrated_final} and \eqref{eq:ODE_corner_layer_integrated_final} (expressed in the original variables) and subtracting their mutual common parts, we obtain a \emph{composite solution} uniformly valid on $\bar{z}\in[0,1]$, to the leading order in $\epsilon$:
\begin{multline}
\bar{u}_{\bar{r}}(\bar{z}) \sim \frac{1}{\beta}\left(\left\{ 1 + 2\beta (3n+2)(1-\nu^2)^2 [{(3+1/n)}]^n (1-\bar{z}) \right\}^{1/(3n+2)} - 1 \right)\\
+ \tilde{C}_4 \re^{-\sqrt[4]{3}\bar{z}/\epsilon} \left[ \sin\left(\sqrt[4]{3}\;\frac{\bar{z}}{\epsilon}\right) + \cos\left(\sqrt[4]{3}\;\frac{\bar{z}}{\epsilon}\right) \right]
- \epsilon \frac{2(1-\nu^2)^2 [{(3+1/n)}]^n}{\sqrt[4]{3}} \re^{-\sqrt[4]{3}(1-\bar{z})/\epsilon}\sin\left(\sqrt[4]{3}\frac{(1-\bar{z})}{\epsilon}\right),
\label{eq:Composite_Solution}
\end{multline}
where the constant $\tilde{C}_4$ is given in Eq.~\eqref{eq:C_4}. 

\begin{remark}
\v{C}anic and Mikeli\'{c} \citep{CM03} discussed the formation of deformation ``boundary layers'' in the context of viscous incompressible flow through a long elastic tube, as the aspect ratio $a/\ell \to 0 ^+$. Their approach was based on \emph{a priori} estimates of the coupled PDEs. Here, we have actually constructed the boundary (and corner) layers explicitly through a matched asymptotic expansion, further showing that the relevant small parameter also involves the tube's thickness: $\epsilon=\sqrt{ta/\ell^2}$. Our result, then, is closer to some of the discussion in textbooks on shell theory, wherein the (dimensional) thickness of boundary layers (near the clamped ends of a Donnell shell subject to uniform internal pressure) is estimated to be $\mathcal{O}(\sqrt{ta})$ \citep[see][Ch.~V]{D90}.
\end{remark}


\section{Results and discussion}
\label{section:results}

\subsection{Deviations from the Hagen--Poiseuille law due to FSI}
\label{section:results1}

\begin{figure}[h]
\centering
\subfloat[Newtonian fluid.]{\includegraphics[width=0.49\linewidth]{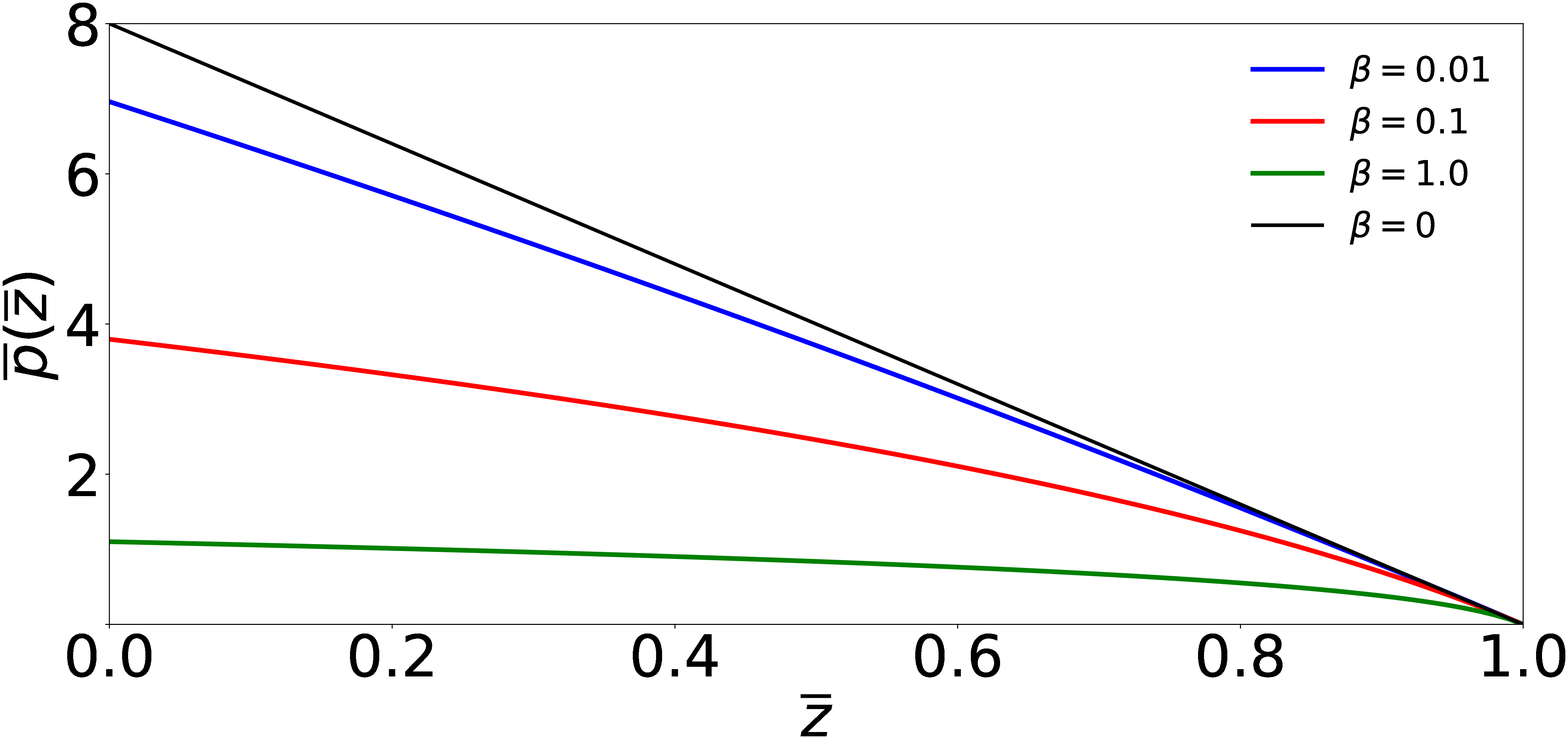}}
\hfill
\subfloat[Shear-thinning fluid.]{\includegraphics[width=0.49\linewidth]{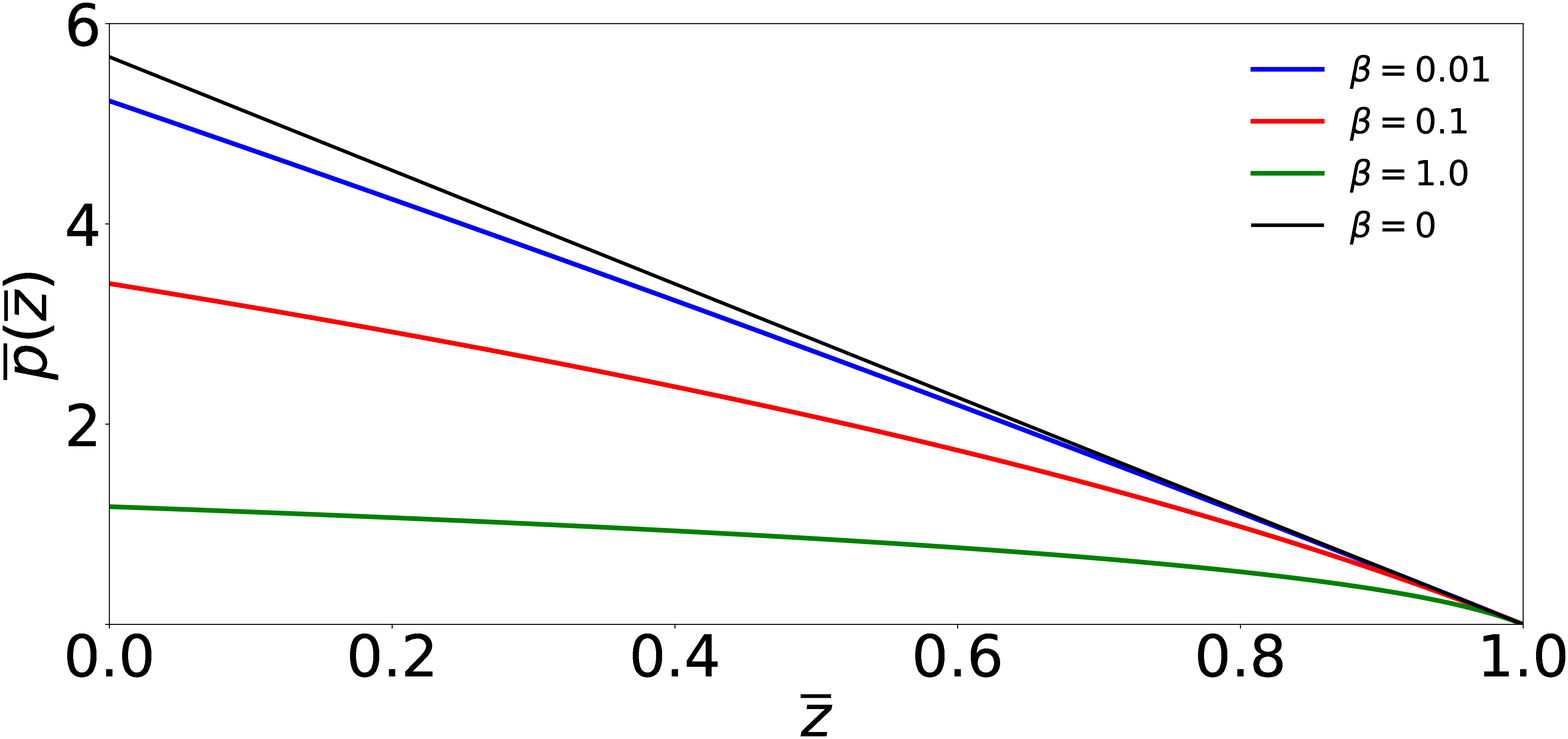}}
\caption{~The dimensionless hydrodynamic pressure $\bar{p}$ in a elastic tube as a function of the dimensionless axial coordinate $\bar{z}$ for different values of the FSI parameter $\beta$ for (a) Newtonian fluid ($n=1$) and (b) shear-thinning fluid ($n=0.7$). Both plots have been produced using Eq.~\eqref{eq:P_vs_Z_Microtube} for an incompressible solid ($\nu=1/2$). Note the different scales on the vertical axes for both the plots. Compliance of the tube reduces the pressure required to maintain steady flow.}
\label{fig:P_Vs_Q_Z}
\end{figure}

Our objective is to quantify the deviation from the Hagen--Poiseuille law caused by FSI in a tube. To this end, we plot the dimensionless pressure $\bar{p}(\bar{z})$ across the tube for different values of the FSI parameter $\beta$ in Fig.~\ref{fig:P_Vs_Q_Z}(a) for Newtonian fluid and in Fig.~\ref{fig:P_Vs_Q_Z}(b) for a shear-thinning fluid. Clearly, ``stronger'' FSI (increasing values of $\beta$) leads to a decrease in the pressure everywhere, but especially near the inlet ($\bar{z}=0$). The decrease in pressure is due to the increase in the flow area, which reduces the resistance to flow. 

\subsection{Comparison between the analytical and numerical solutions for the flow-induced deformation}
\label{section:results2}

Our results in \S\ref{section:coupling}, suggest the following possible ways for solving the coupled problem of flow and deformation in an elastic tube: (i) using leading-order-in-thickness membrane theory (\S\ref{sec:leading_order_fsi}), (ii) using a matched asymptotic expansion for beyond-leading-order-in-thickness (i.e., Donnell shell) theory to capture bending and clamping (\S\ref{sec:beyond_leading_order_fsi}), and (iii) by numerical integration of the nonlinear TPBVP for the displacement given by Eqs.~\eqref{eq:coupled_ODE_non_dim} and \eqref{eq:coupledBC}. We may conceptualize these approaches as a hierarchy: the leading-order perturbative solution is a less accurate version of the solution obtained by the matched asymptotic expansion, which in itself is a less accurate version of the solution found by solving the TPBVP numerically.

\begin{figure}
\centering
\subfloat[]{\includegraphics[width=0.49\linewidth]{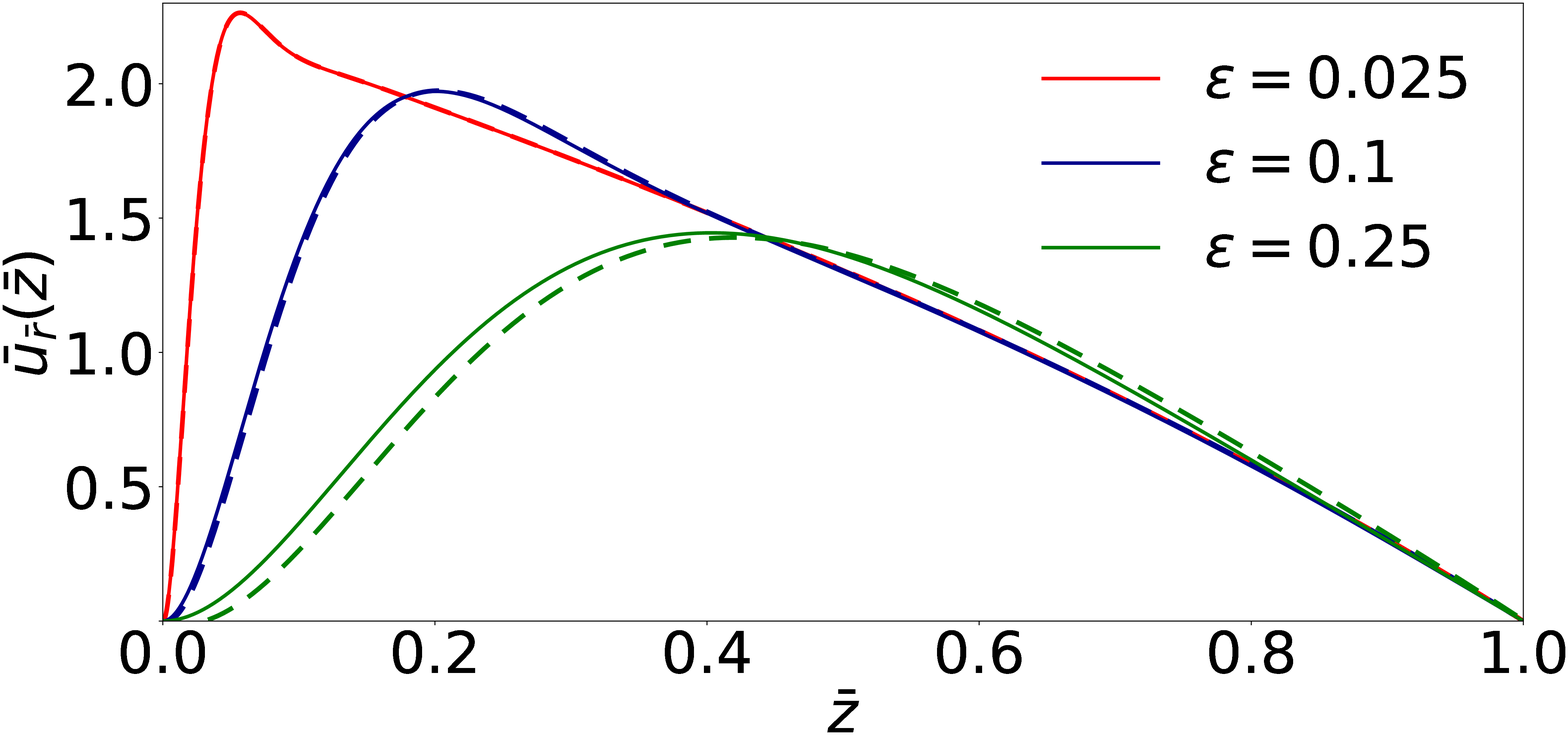}}
\hfill
\subfloat[]{\includegraphics[width=0.49\linewidth]{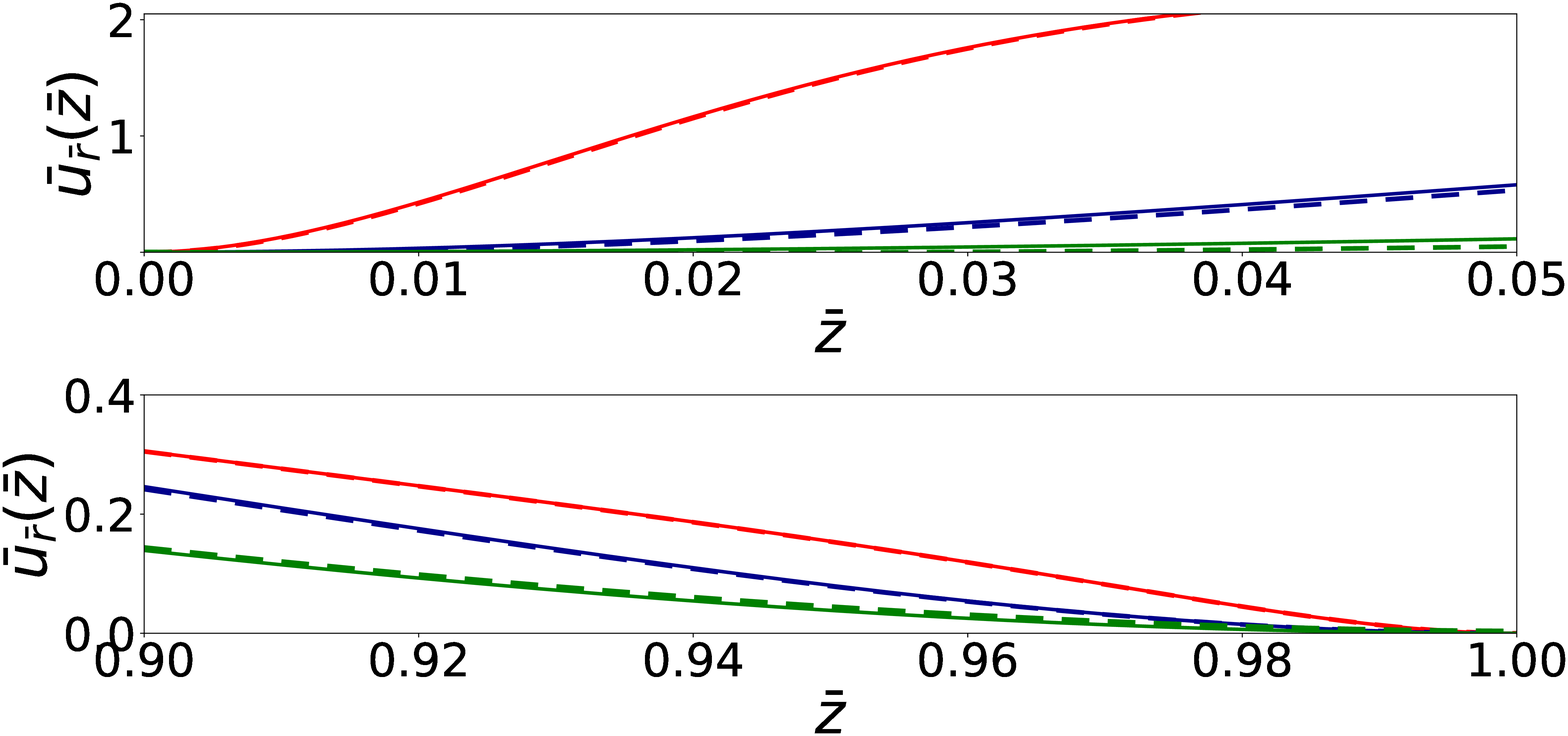}}
\caption{~The dimensionless radial displacement $\bar{u}_{\bar{r}}(\bar{z})$ as a function of the dimensionless axial position $\bar{z}$ in the elastic micro for $\beta=0.11$, $n=0.7$, and $\nu=1/2$. The solid curves are the numerical solution of TPBVP given by Eqs.~\eqref{eq:coupled_ODE_non_dim} and \eqref{eq:coupledBC}, while dashed curves are the matched asymptotic solution from Eq.~\eqref{eq:Composite_Solution}. Panel (a) shows the displacement over the whole tube, while panels (b)  and (c) show zoom-ins near the clamped ends. The matched asymptotic solution is highly accurate, capturing the displacement overshoot near the inlet as $\epsilon\to0^+$.}
\label{fig:MT_Deflection_Vs_Z_Matched}
\end{figure}

Let us now compare the deformation profile obtained via matched asymptotic expansion, i.e., Eq.~\eqref{eq:Composite_Solution} to the numerical solution of the original nonlinear TPBVP. The latter profile is obtained using the {\tt solve\_bvp} method in Python's SciPy module \citep{SciPy} to solve the TPBVP numerically. Figure \ref{fig:MT_Deflection_Vs_Z_Matched} shows the results of such a comparison for different values of the small parameter $\epsilon=\sqrt{ta/\ell^2}$ but fixed $\beta$, $n$, and $\nu$. There is very good agreement between the composite solution obtained via a matched asymptotic expansion and the numerical solution of the nonlinear TPBVP. As expected, the error in the composite solution increases with $\epsilon$, especially in the corner layer at $\bar{z}=1$. Nevertheless, the asymptotic expression is clearly very accurate.

In addition, observe that the radial displacement profile exhibits an overshoot near the inlet due to clamping. A similar profile has also been reported by Heil and Pedley~\citep{HP95} in their numerical study of large-deformation, small-strain FSI in a \emph{collapsible} tube, which they modeled using Poiseuille's law and geometrically nonlinear shell theory (accounting for axial pre-stretch). However, since Heil and Pedley~\citep{HP95} modelled buckling of collapsible tubes, as opposed to the inflated tubes studied herein, the deformation profile actually exhibits an undershoot (compare \citep[Fig.~6]{HP95} with Fig.~\ref{fig:MT_Deflection_Vs_Z_Matched} above). Perhaps more importantly, even though Heil and Pedley~\citep{HP95} accounted for bending stiffness of the tube, their analysis does not yield an analytical solution to the deformation profile, i.e., a  counterpart to Eq.~\eqref{eq:Composite_Solution} derived above. We also observe that the deformation profile of a beam due to quasi-static gravity-driven spreading of a viscous fluid underneath it exhibits this overshoot near the edge [see the inset of \citep[Fig. 2(a)(ii)]{BN18}], as should be expected from the structure of solutions to Eq.~\eqref{eq:Cylinder_Structure}.

\subsection{Comparison between theory and direct numerical simulations: Flow and deformation}
\label{sec:theory_vs_simulat}

To ascertain the validity of the theory developed in this paper, we now compare our theoretical results against 3D direct numerical simulations (DNS) of coupled flow and deformation in an elastic tube. To this end, we choose an illustrative set of physical and geometric parameters, given in Table~\ref{Table:MicrotubeGeometry}. The tube is assumed to be made of {elastin}, which is a highly elastic protein found in all vertebrates and is major constituent of arteries \citep{SGF77}. Here, along the lines of the work in \citep{SGF77,MSY15,E09}, elastin is modeled as an isotropic linearly elastic solid with a constant Young's modulus of $E=0.5$ MPa and a Poisson ratio of $\nu=0.499$ (i.e., a nearly incompressible material). The dimensions have been chosen to ensure that the assumptions of shallowness and slenderness are satisfied.  Specifically, the thickness-to-radius ratio of the tube is fixed at $t/a=0.1$, consistent with the hemodynamics literature \cite[p.~60]{VON11}. The radius-to-length ratio is chosen to be sufficiently smaller, $a/\ell=0.025 \ll t/a$, and the length $\ell = 3.2$ mm is chosen to be similar to microchannel studies \cite[see, e.g.,][]{SC18}, from which the values of $a$ and $t$ in Table~\ref{Table:MicrotubeGeometry} follow. 

The generalized Newtonian fluid inside the tube is assumed to be human blood. Blood rheology is a topic of active research, as the rheological properties of blood depend on various factors such as a patient's age, health, concentration of plasma, etc.~\citep{HKP99,C05}. Here, for the sake of simplicity, without sacrificing any physics, and to validate our theory for both Newtonian and non-Newtonian rheology, blood plasma is chosen as our example of a Newtonian fluid with constant shear viscosity of $\mu=0.0012$ Pa$\cdot$s (i.e., $m=\mu$ and $n=1$) \citep{F97}, while whole blood is chosen as our example of a shear-thinning fluid with a power-law index of $n=0.7$ and a consistency index of $m=0.0185$ Pa$\cdot$s$^n$ \citep{HKP99}. In both cases, a density of $\rho = 1060$ kg/m$^3$ is used, which is within the range for both blood plasma and whole blood \cite[see, e.g.,][Table 2.1.1]{K06_book}. Simulations were carried out for flow rates up to $q=2.00$ mL/min, which corresponds to a maximum $Re \approx 150$ [$\Rightarrow \lambda Re \approx 3.75$, which is reasonably small within the lubrication approximation and well below the onset of unsteady effects closer to $Re\approx 300$ \citep{BT06}] and a maximum FSI parameter value of $\beta \approx 0.12 \ll 1$.

\begin{table}
\centering
\begin{tabular}{lllllllllll}
  \hline
  $a$ (mm) & $\ell$ (mm) & $a/\ell$ (--) & $t$ (mm) & $t/a$ (--) & $E$ (MPa) & $\nu$ (--) & $\mu$ (Pa$\cdot$s) & $n$ (--) & $m$ (Pa$\cdot$s$^n$) & $\varrho$ (kg/m$^3$)\\
  \hline
  \hline
  0.08 & 3.2 & 0.025 & 0.008 & 0.1 & 0.5 & 0.499 & 0.0012 & 0.7 & 0.0185 & $1\,060$\\
  \hline
\end{tabular}
\caption{~Geometric and material properties for a sample tube FSI problem.}
\label{Table:MicrotubeGeometry}
\end{table}

For our computational approach, we employ a segregated solution strategy, as opposed to a monolithic one \citep[see, e.g.,][]{BTT13}. That is, the solid (resp.\ fluid) problem is solved independently of the fluid (resp.\ solid) problem, each on its own computational domain. The displacements (resp.\ forces) from the solid (resp.\ fluid) domain are then transferred to the fluid (resp.\ solid) domain via a surface traction boundary condition. Based on previous successful computational microscale FSI studies \citep{CPFY12,SC18,ADC18}, we have used the commercial computer-aided engineering (CAE) software from ANSYS~\citep{ANSYS2} to perform such two-way coupled FSI DNS via a segregated approach. The domain geometry was created in ANSYS SpaceClaim, and the computational grid was generated in ANSYS ICEM CFD. The steady incompressible mass and momentum equations for the power-law fluid were solved using ANSYS Fluent as the computational fluid dynamics (CFD) solver based on the finite volume method. The structural mechanics solver, based on the finite element method (FEM), under ANSYS Mechanical was employed  for the structural problem, which solved the static force-equilibrium equations for a linearly elastic isotropic solid with geometrically nonlinear strains. In the static structural module, the option of ``large deformations'' had been turned on. This feature allows the distinction between the deformed and undeformed coordinates to be retained. Additionally, this feature means that that the Henky strain and Cauchy stress are used as the strain and stress measures, respectively. Consequently, the stiffness matrix in the resultant finite element method (FEM) formulation is not constant, and it is a function of displacements, leading to a nonlinear algebraic problem. However, there is no material nonlinearity in this problem, and the relationship between the stress and strain tensors is described by linear elasticity with two material parameters ($E$ and $\nu$). The exchange of forces and displacements along the inner surface of the tube was achieved by declaring the surface as an ``FSI interface.'' A nonlinear iterative procedure transfers the loads and displacements incrementally until convergence is reached, ensuring two-way coupling. Most importantly, beyond assuming a steady state, this DNS approach \emph{does not} make any of the approximations that the theory does, e.g., lubrication, Donnell shell, and the various smallness assumptions.

\subsubsection{Fluid mechanics benchmark}

\begin{figure}
\subfloat[Newtonian fluid.]{\includegraphics[width=0.45\linewidth]{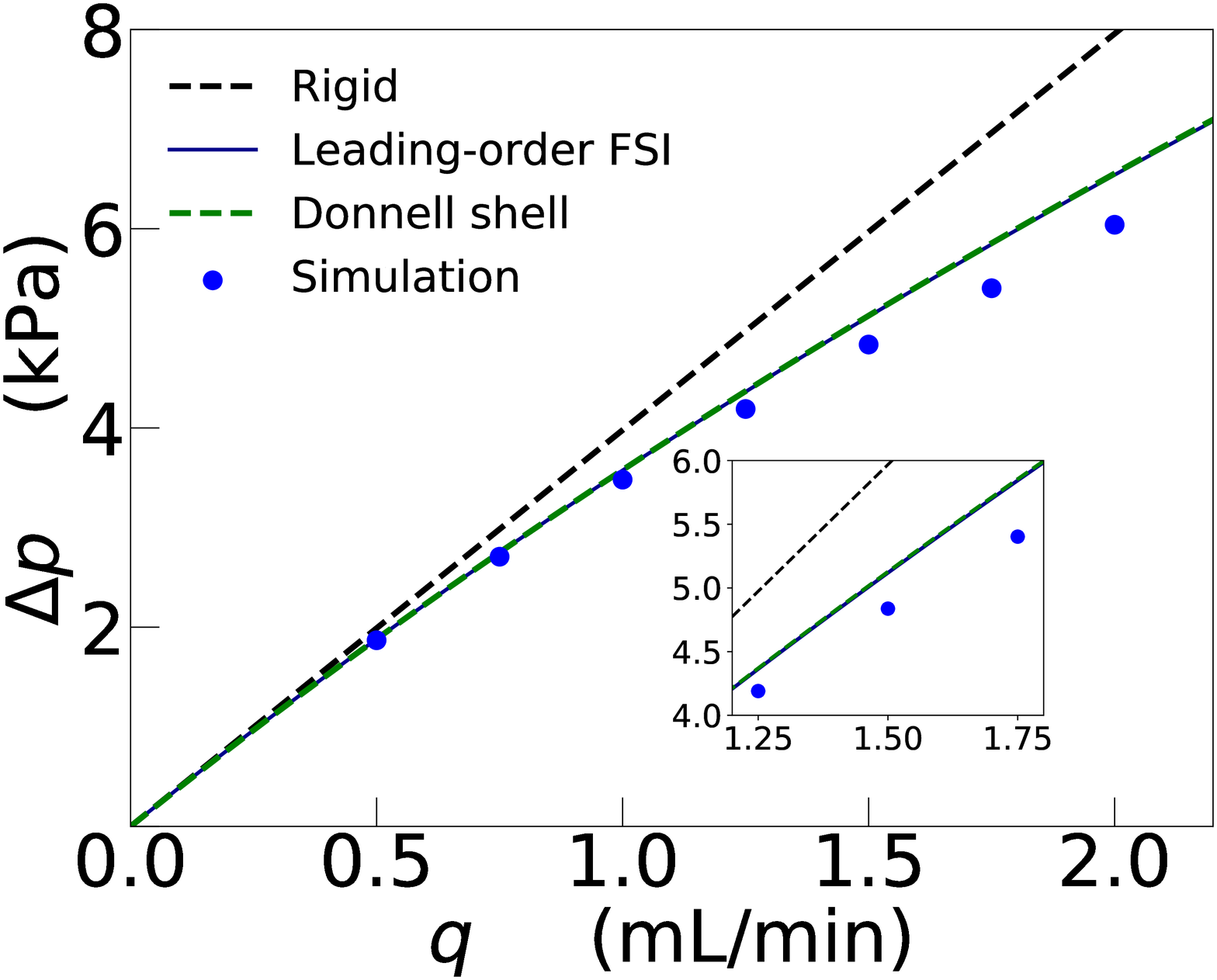}}
\hfill
\subfloat[Shear-thinning fluid.]{\includegraphics[width=0.45\linewidth]{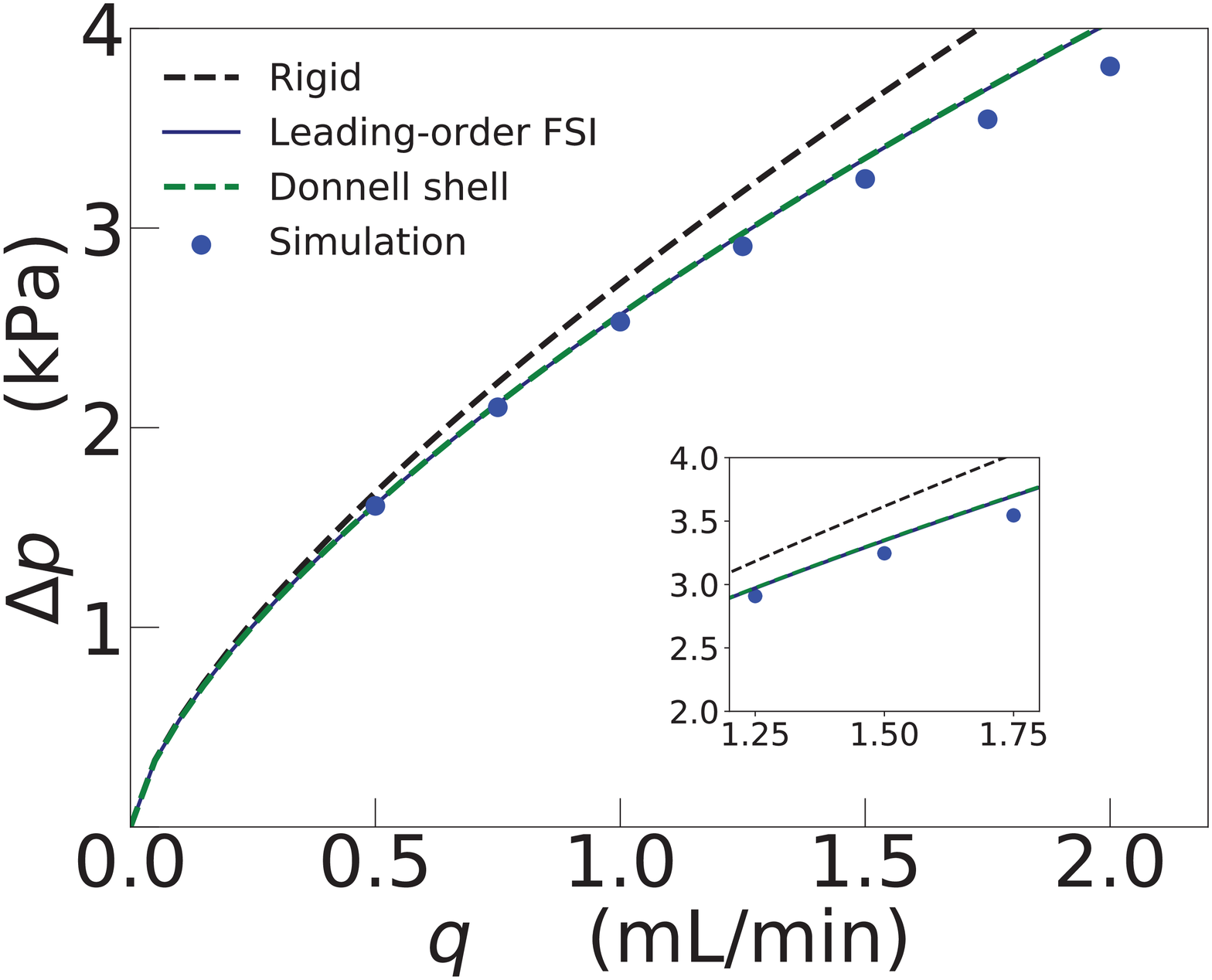}}
\caption{~Full pressure drop $\Delta p = p(0)$ vs.\ flow rate $q$ in a deformable tube. Fluid--structure interaction causes the pressure drop to decrease in the deformable tube compared to the rigid one. The perturbative analysis developed herein, culminating in Eqs.~\eqref{eq:P_vs_Z_Microtube} [power-law, (b)] and \eqref{eq:P_vs_Z_Microtube_Newtonian} [Newtonian, (a)] captures the latter effect quite accurately, when compared to 3D direct numerical simulations. Note the different scales on the vertical axes for both the plots.}
\label{fig:MT_Simulation_Delta_P_Vs_Q}
\end{figure}

First, we benchmark the $q$--$\Delta p$ relationships predicted by our mathematical models: the leading-order FSI from the membrane theory [Eq.~\eqref{eq:DP_vs_Q_Microtube}] and the Donnell shell FSI [Eqs.~\eqref{eq:coupled_ODE_non_dim} and \eqref{eq:coupledBC}]. The dimensional full pressure drop $\Delta p$ as a function of the dimensional flow rate $q$ is shown in Fig.~\ref{fig:MT_Simulation_Delta_P_Vs_Q} for (a) a Newtonian fluid (blood plasma) and (b) a shear-thinning fluid (whole blood). There is good agreement between theory and simulation, particularly for the smaller flow rates. The maximum relative error, over the shown range of $q$, is $\approx 10$\% for the Newtonian fluid and $\approx 5$\% for the shear-thinning fluid. The higher relative error for the same flow rate for Newtonian fluid flow is attributed to the larger pressure (note the different vertical axis scales in the two plots). At larger $q$, small but systematic differences emerge between theory and simulation because, at these flow rates, the deformation of the tube starts to exhibit significant strains, which are beyond the applicability of the linear shell theory employed herein. We also observe that there is hardly any perceptible difference in $\Delta p$ predicted by the membrane and Donnell shell theories, which shows that, indeed, bending has a negligible effect on the total pressure drop.

{To justify the use of the power-law rheological model in the fluid mechanics problem, we post-processed our simulation results, and we verified that the maximum shear stress in the flow is much larger than the yield stress of the Casson model for blood. For example, for the flow rate of $1$ mL/min, the maximum wall shear stress on the tube was found to be $35.14$ Pa, which is two orders of magnitude higher than the $0.05$ Pa yield stress of the Casson model for blood \cite{BS70}.}

Next, it is important to evaluate the theory developed herein in the context of the classical results, namely the law of Laplace \citep{Canic2006,CGM07} and the model proposed by Fung in his \textit{Biomechanics} textbook \cite[\S3.4]{F97}, which is often quoted in newer texts on biofluid mechanics \cite[pp.~e25--e27]{A16}. Fung's model and that of Laplace take a large-deformation approach, writing the stress equilibrium equations in the \emph{deformed} configuration of the tube, which requires reconsidering the structural mechanics calculation from \S\ref{sec:membrane}. In order not to belabor the discussion in this section, we have relegated the large-deformation re-derivation of our main mathematical result [i.e., Eq.~\eqref{eq:P_vs_Z_Microtube}] to Appendix~\ref{Appendix:Fung}.

\subsubsection{Structural mechanics benchmark}

\begin{figure}
\centering
\subfloat[Newtonian fluid.]{\includegraphics[width=0.45\linewidth]{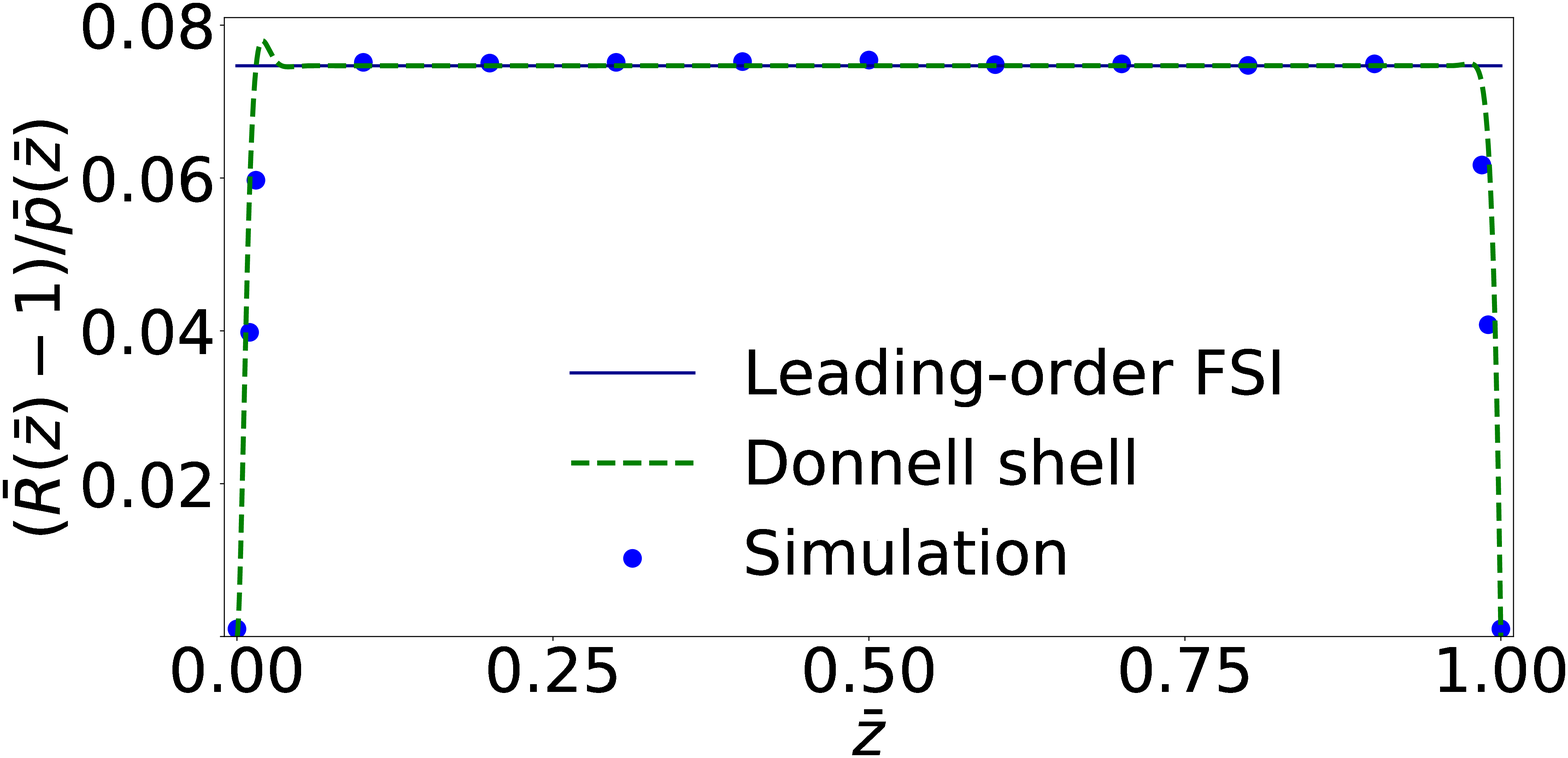}}
\hfill
\subfloat[Shear-thinning fluid.]{\includegraphics[width=0.45\linewidth]{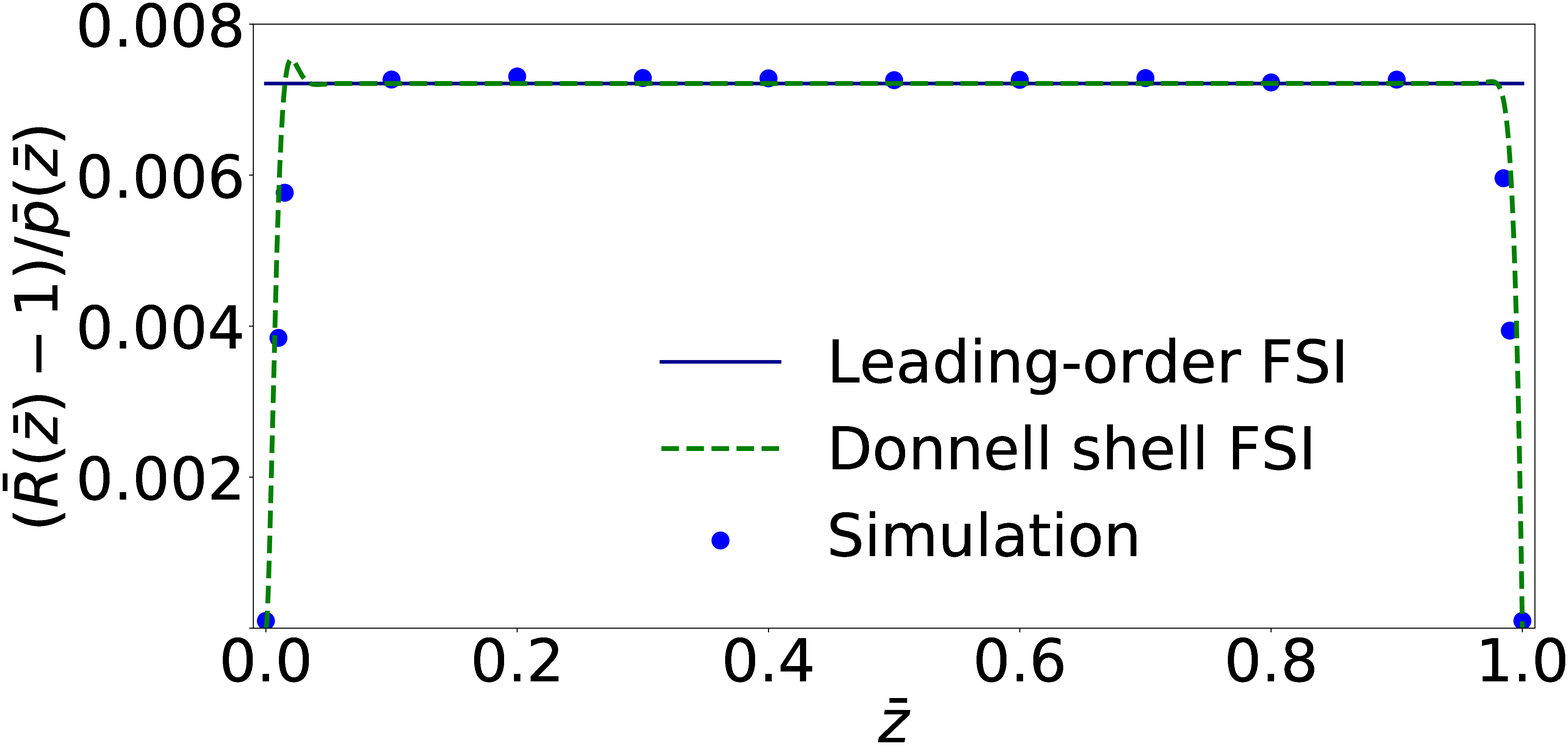}}
\caption{~Ratio of the dimensionless radial deformation $\bar{R}-1$ to the hydrodynamic  pressure $\bar{p}$, as a function of the axial position $\bar{z}$ in the tube. The leading-order FSI theory is given by Eq.~\eqref{nondimensional_reduced_deflection_microtube} (solid), the Donnell shell FSI theory is the numerical solution of the TPBVP given by Eqs.~\eqref{eq:coupled_ODE_non_dim} and \eqref{eq:coupledBC} (dashed), the simulation results are from ANSYS (symbols). Note the different vertical scale of these plots. For both the plots, the results of the simulations correspond to a flow rate of $q = 1$ mL/min.}
\label{fig:MT_Deflection_Vs_Z}
\end{figure}

Having compared and validated the theoretical prediction for the hydrodynamics portion of the FSI problem, we now shift our focus to the solid domain. In Fig.~\ref{fig:MT_Deflection_Vs_Z}, we plot the ratio of dimensionless radial tube deformation, $\bar{R}(\bar{z})-1$, to the dimensionless pressure $\bar{p}(\bar{z})$ along the tube's length. Again, the results have been shown for both our chosen (a) Newtonian and (b) shear-thinning fluid. Equations~\eqref{eq:R1_nd} and \eqref{nondimensional_reduced_deflection_microtube} predict this ratio to be $[\bar{R}(\bar{z})-1]/\bar{p}(\bar{z})=\beta(1-\nu^2)$ (interestingly, a constant independent of $\bar{z}$), to the leading order in $\epsilon=\sqrt{ta/\ell^2}$. For the beyond-leading-order analysis, we integrated the fifth-order nonlinear BVP from Donnell's shell theory, i.e., Eq.~\eqref{eq:coupled_ODE_non_dim} subject to Eqs.~\eqref{eq:coupledBC}, using the TPBVP solver method in Python's SciPy module \citep{SciPy}. (We could have also plotted the matched asymptotic solution to Eq.~\eqref{eq:coupled_ODE_non_dim}, i.e.,  Eq.~\eqref{eq:Composite_Solution}, but it would be indistinguishable for this value of $\epsilon=6.25\times 10^{-5}$ based on Table~\ref{Table:MicrotubeGeometry}.) 

Figure~\ref{fig:MT_Deflection_Vs_Z} shows good agreement between the results of DNS and the two proposed mathematical models (i.e., leading-order FSI and the Donnell shell FSI). As predicted in \S\ref{sec:beyond_leading_order_fsi} there are extremely narrow regions (boundary layers) near the inlet ($\bar{z}=0$) and outlet ($\bar{z}=1$) planes in which the full ODE solution deviates from the leading-order perturbative solution. The thinness of these boundary layers, in comparison with the tube's length, leads us to conclude that almost the entire tube, except a sliver near each end, is in a membrane state with negligible bending.

To verify our assumption about negligible (no) axial displacement, which was used to arrive at Eq.~\eqref{eq:equal_stress_uz_zero}, we post-processed the axial $\sigma_{zz}$ and hoop $\sigma_{\theta\theta}$ stresses from our ANSYS simulations. For a Newtonian fluid, the ratio of the hoop stress to the axial stress at the middle of the tube, where the effect of clamping at the edges and the resultant bending is negligible, is $(\sigma_{\theta\theta}/\sigma_{zz})|_{\bar{z}=0.5} \approx 2.20$ for $q =0.5$ mL/min and $(\sigma_{\theta\theta}/\sigma_{zz})|_{\bar{z}=0.5} \approx 2.45$ for $q = 2$ mL/min. These two values, for the smallest and largest flow rates considered, can be compared to the theoretical value of $1/\nu=2$ [from Eq.~\eqref{eq:equal_stress_uz_zero}], leading to a relative error of $10\%$ for $q =0.5$ mL/min and $22\%$ for $q =0.5$ mL/min. Similarly, the ratio of axial deformation to radial deformation at the middle of the tube is $\approx 0.004$ for $q=0.5$ mL/min and $\approx 0.2$ for $q =2$ mL/min. While nonzero, these errors are small enough to justify having employed Eq.~\eqref{eq:equal_stress_uz_zero} in the mathematical analysis above.

\subsection{Region of validity in the parameter space}
\label{section:results4}

The proposed theory of steady non-Newtonian FSI in slender elastic tubes hinges upon on a set of intertwined assumptions:
\begin{enumerate}
\item ${t}/{a} \ll 1$: This requirement allows us to use thin-shell theory. 
\item ${\mathcal{U}_c}/{a} \ll 1$: This requirement represents the small-strain assumption of the shell theory.
\item ${a}/{\ell} \ll 1$: This slender-geometry requirement allows us to simplify both the fluid mechanics and the structural mechanics problems. This requirement also ensures the rotation of a shell element is negligible. 
\item ${\mathcal{U}_c}/t \ll 1$: This requirement allows us to refer the analysis to the undeformed (Eulerian) coordinates and also restricts the theory to small deformations. 
\end{enumerate}
A natural ordering of the length scales associated with FSI in a tube thus follows: 
\begin{equation}
\mathcal{U}_c \ll t \ll a \ll \ell,
\label{eq:scale_separation_final}
\end{equation}
which must hold for the present linear, small-deformation FSI theory to apply. Importantly, our DNS results form \S\ref{section:results2} show that this regime is accessible under realistic flow conditions.


\section{Conclusions and outlook}
\label{section:conclusion}

In this paper, we formulated a theory of the low Reynolds number axisymmetric fluid--structure interaction (FSI) between a generalized Newtonian fluid and an elastic tube enclosing the flow. Specifically, we derived an analytical relation between the pressure drop across the tube and the imposed steady flow rate through it, taking into account both the fluid's shear-dependent viscosity (such as, e.g., whole blood) and the compliance of the conduit (such as, e.g., a blood vessel). Although physiological flows occur across a range of flow regimes (Reynolds numbers) \citep{P80,G94,GJ04,HH11}, previous research has focused on moderate-to-high Reynolds number phenomena, including collapse of the vessel, unlike the present context. The proposed theory is also applicable to problems in microfluidics, wherein soft (e.g., PDMS-based) microchannels of circular cross-section and microtubes are now manufactured and FSI becomes relevant \citep{PCK15,RCDC18}.

Under the lubrication approximation for low Reynolds number flow, we showed how to analytically couple a unidirectional flow field to an appropriate linear shell theory. This led us to a fifth-order nonlinear ordinary differential equation (ODE), namely Eq.~\eqref{eq:coupled_ODE_non_dim} for the radial displacement, which fully describes the FSI. In a perturbative sense, we showed that, at the leading order in slenderness and shallowness of the tube, the shell theory reduces to membrane theory and a \emph{linear} relationship, given by Eq.~\eqref{nondimensional_reduced_deflection_microtube}, emerges between the local radial deformation and the local hydrodynamic pressure at a given cross-section. Then, a dimensionless ``generalized Poiseuille law'' was obtained in  Eq.~\eqref{eq:DP_vs_Q_Microtube}, which \emph{explicitly} gives the pressure drop in terms of the solid and fluid properties. This relationship rationalizes and updates certain arguments found in textbook discussions of physiological flow \citep{F97,A16}. Specifically, Eq.~\eqref{eq:DP_vs_Q_Microtube} can be put into dimensional form to yield the total pressure drop $\Delta p$ due to flow of a generalized Newtonian (power-law) fluid in a slender deformable tube:
\begin{equation}
\Delta p = \frac{Et}{(1-\nu^2)a} \left\{\left[ 1+ 2(3+1/n)^n(3n+2)(1-\nu^2)\left(\frac{m \ell }{Et}\right)\left(\frac{q}{\pi a^3}\right)^n\right]^{1/(3n+2)}-1\right\}.
\label{eq:P_vs_Z_Microtube_Newtonian_dim}
\end{equation}
The most important observation is that, due to FSI, Eq.~\eqref{eq:P_vs_Z_Microtube_Newtonian_dim} is \emph{nonlinear} in $q$, even for a Newtonian fluid ($n=1$), in contrast to the Hagen--Poiseuille law.

Furthermore, we showed that a boundary layer (at the inlet) and a corner layer (at the outlet) of the tube are required to enforce the clamping conditions on the structure. Using the method of matched asymptotics, we obtained a uniformly valid (closed-form) expression, given by Eq.~\eqref{eq:Composite_Solution}, for the deformation profile. The ability to solve for the tube's deformation as a function of the axial coordinate via a matched-asymptotics calculation is in contrast with the case of low Reynolds number FSI in a microchannel of rectangular cross-section \citep{CCSS17,ADC18} (see also \S\ref{sec:plate_vs_shell}), but similar to planar low Reynolds number flow under an elastic beam \citep{HBDB14}.

To ascertain the validity of our mathematical results, we carried out two-way-coupled 3D simulations using the commercial computer-aided engineering suite by ANSYS~\citep{ANSYS2}. We showed that good agreement can be obtained between predictions from the theory (for both the pressure drop and the radial deformation) and the corresponding direct numerical simulations. Then, we specified, through  Eq.~\eqref{eq:scale_separation_final}, the region in the physical and geometric parameter space in which our FSI theory applies. 

In future work, the FSI theory developed herein can be extended to incorporate further physical effects that arise in microscale fluid mechanics. For example, the material composing the tube may not be only elastic but also porous (i.e., \emph{poroelastic}) \citep{AM17}. It may also be worthwhile to consider microflows of gases in elastic tubes, which necessitates accounting for compressibility of the fluid \citep{EJG18,MY18} and, possibly, wall slip \citep{SS12,CS18}. Another potential avenue for future research stems from the fact that many soft biological tissues are \emph{hyperelastic}. In this case, the stress--strain relationship is obtained from extremizing a prescribed strain-energy functional (see, e.g., \cite[Ch.~8 and 9]{F93}). With such an elastic response in hand, the tube FSI problem considered herein can be generalized, along the lines of \citep{AC19a}, wherein the leading-order (in the present terminology) hyperelastic FSI was studied. 

\begin{figure}[hb!]
  \centering
  \includegraphics[width=0.65\linewidth]{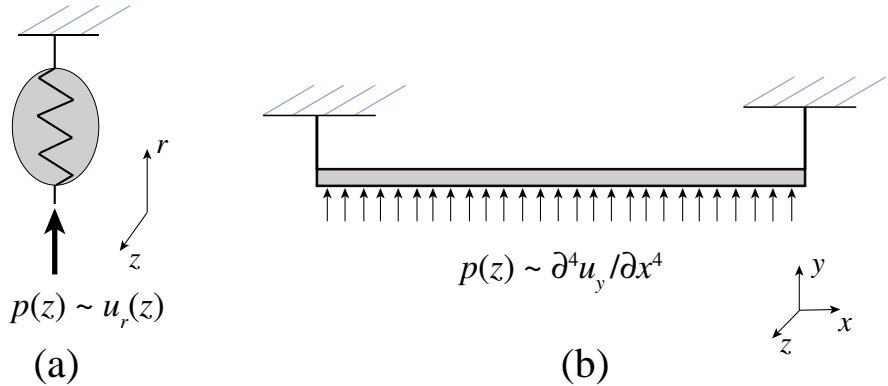}
  \caption{~Shells versus plates in viscous FSI. (a) A slice of a soft microtube represented as a Hookean spring. (b) A slice of a microchannel's soft top wall represented as an Euler beam. Under the perturbation theory, each slice is uniformly loaded by the hydrodynamic pressure $p$ at flow-wise cross-section $z$.}
\label{fig:schematic}
\end{figure}

\subsection{Comparison between viscous FSI in a rectangular microchannel and an axisymmetric microtube}
\label{sec:plate_vs_shell}

As mentioned previously, the deformation of any infinitesimal transverse slice in the streamwise direction of the elastic tube considered herein is akin to a Hookean spring because of the direct \textit{linear} proportionality between the deformation and the pressure expressed by Eq.~\eqref{nondimensional_reduced_deflection_microtube}. This deformation is independent of neighboring slices, just as for a slender microchannel with a soft top wall \citep{CCSS17,SC18,ADC18}. The mathematical reason for this decoupling is the same for both geometries: the streamwise length scale of the microtube/channel is much larger than its cross-sectional dimension. However, the two cases should also be contrasted: for a long and shallow microchannel, each slice of the top wall acts like an Euler beam in the cross-sectional direction (due to clamping at the sidewalls); while, for a slender microtube, each slice acts as a Hookean spring (due to axisymmetric deformation), as illustrated schematically in Fig.~\ref{fig:schematic}. The maximum cross-sectional deformation of the microtube, from the dimensional versions of Eqs.~\eqref{eq:R1_nd} and \eqref{nondimensional_reduced_deflection_microtube},  is $u_\mathrm{max}(z) = (1-\nu^2)a^2p(z)/(Et)$, while it is $u_\mathrm{max}(z) = (1/2)(1-\nu^2)a^4p(z)/(Et^3)$ for a microchannel with equivalent width $w=2a$ \citep{CCSS17}. Consequently, a microtube is more compliant (and thus deforms more than a similar microchannel), with all other conditions being the same, if $t < a/\sqrt{2}$. This condition is satisfied in the distinguished limit considered herein: $t/a\ll1$.




\section*{Acknowledgements}
We thank T.\ C.\ Shidhore for assistance with the initial forays into two-way ANSYS FSI simulations and advice related to the latter, and X.\ Wang for a careful reading of the manuscript. This research was supported by the US National Science Foundation under grant No.\ CBET-1705637.



\bibliography{Mendeley_bibliography}


\appendix

\renewcommand\thetable{A\arabic{table}}

\section{Grid-independence study}
\label{Appendix:ANSYS}

To ensure that the numerical solutions computed are independent of the grid choice, we performed a grid-inde\-pend\-ence (convergence) study. Two grids each were defined for the fluid and the solid solvers, thus bringing the total number of grid arrangements to four. The details of the grids are shown in Table~\ref{Table:GridIndependence}. For the fluid grid, grid 2 was generated by scaling the number of edge divisions across the model in grid 1 by a factor of 2. Similarly, grid 2 for the FEM grid was generated by increasing the number of divisions on the lateral surfaces from $500$ to $800$. The simulations were performed for the fluid and solid models on all four grids described above, under the conditions in Table~\ref{Table:MicrotubeGeometry}. However, for the grid refinement study, the simulations were carried out only for a single flow rate of $q=1$ mL/min. 

\begin{table}[h]
\centering
  \begin{tabular}{l|cccc}
    \hline
	{} & fluid grid 1 & fluid grid 2 & solid grid 1 & solid grid 2 \\
	\hline
	\hline
	Number of nodes & $1\,348\,768$ & $10\,626\,967$ & $1\,090\,584$ & $1\,743\,984$ \\ 
	Number of elements & $1\,387\,365$ & $10\,467\,576$ & $198\,000$ & $316\,800$\\
	\hline
  \end{tabular}
\caption{~Details for the four grids used for the grid-convergence study.}
\label{Table:GridIndependence}
\end{table}

The results of the grid convergence study are shown in Table~\ref{Table:GridIndependence2} for the  displacement at the midsection of the tube (averaged over the circumference) and the pressure drop over the length of the microtube. The insignificant variation of these values across grid combinations shows that our simulation results, which were computed on the combination fluid grid 1 and solid grid 1, are indeed grid independent and accurate.

\begin{table}[h]
\centering
  \begin{tabular}{l|cccc}
    \hline
	{fluid grid/solid grid} &  1/1 & 2/1 & 1/2 & 2/2\\
	\hline
	\hline
	$\langle u_r\rangle$ \% difference &$0.2$ &$-0.4$ & $-0.6$ & --\\ 
	$\Delta p$ \% difference &$-0.08$ & $0.08$ & $0.04$ & --\\
	\hline
  \end{tabular}
\caption{~Grid-independence (convergence) study for the ANSYS simulations, using the circumferentially averaged displacement $\langle u_r\rangle$ at the tube wall's midsection and the full pressure drop $\Delta p$ as the metrics. The percent difference is computed with respect to the reference values from the simulation on the combination of fluid grid 2 and solid grid 2.}
\label{Table:GridIndependence2}
\end{table}


\renewcommand\thefigure{B\arabic{figure}}

\section{Large-deformation formulation and connections to the law of Laplace and Fung's model}
\label{Appendix:Fung}

In \S\ref{section:deformation}, we termed the proposed FSI theory as \textit{small deformation} due to our assumptions on the structural mechanics aspect of the FSI. We used small-deformation classical shell theories, which assume that the (radial) deformation is considerably smaller than the (smallest) characteristic dimension of the tube, i.e., $\mathcal{U}_c/t \ll 1$. This assumption ensures the equivalence between the deformed and undeformed coordinates (i.e., between the Langrangian and Eulerian frames). Then, the equations of static equilibrium, which are strictly valid only when written in terms of deformed coordinates, can be written in terms of the undeformed coordinates, as done above and in classical shell theory. However, some prior work in the literature has used a mixture of frames of references by posing the equilibrium equations of a small-deformation shell theory in the deformed configuration, while using the undeformed coordinates in the subsequent calculation of deformations from strains. Thus, in this appendix, we offer a critical discussion of this issue. The resulting theory may be termed ``large-deformation'' theory.
 
The membrane theory of \S\ref{sec:membrane}, when referred to the tube's deformed coordinates, leads us to reformulate Eqs.~\eqref{eq:Ntheta_p} and \eqref{eq:hoop_stress} as 
\begin{equation}
  N_{\theta} = R(z)p(z)\quad\Rightarrow\quad \sigma_{\theta\theta} = \frac{R(z)}{t} p(z).
\end{equation}
The assumption of no axial displacement leads to Eq.~\eqref{eq:equal_stress_uz_zero}, which still applies and whence the hoop strain (from the linear elastic law) is $\varepsilon_{\theta\theta} = (1-\nu^2)\sigma_{\theta\theta}/E$. Then, the radial deformation, under the assumption of axisymmetry, is
\begin{equation}
\label{eq:Laplace_Law}
u_r(z) = \varepsilon_{\theta \theta} a =  (1-\nu^2)\frac{R(z)a}{Et}p(z).
\end{equation}
Equation~\eqref{eq:Laplace_Law} is the so-called \emph{law of Laplace}, which relates the pressure at a given cross-section to its radius in the \emph{deformed} state \citep{Canic2006,CGM07}. This result, in mixing frames of reference without rigorous motivation for doing so, is not without criticism in the biomechanics literature \citep{CB15}. Furthermore, observe that $u_r$ appears on both sides of Eq.~\eqref{eq:Laplace_Law} because $R(z) = a + u_r(z)$.

Using the dimensionless variables from Eq.~\eqref{eq:nd_vars_tube2}, Eq.~\eqref{eq:Laplace_Law} becomes
\begin{equation}
\label{eq:Radial_Deformation_Membrane_Undeformed}
\mathcal{U}_c\bar{u}_{\bar{r}}(\bar{z}) = (1-\nu^2)\frac{(a+\mathcal{U}_c \bar{u}_{\bar{r}})a}{Et}\mathcal{P}_c\bar{p}(\bar{z}).
\end{equation}
Now, for large deformations, the appropriate scale is $\mathcal{U}_c = t$, which ensures consistency with the results above  by keeping the FSI parameter defined as $\beta = a\mathcal{P}_c /(E t)$. Thus, solving for $\bar{u}_{\bar{r}}$ from Eq.~\eqref{eq:Radial_Deformation_Membrane_Undeformed}, we obtain the pressure--deformation relationship:
\begin{equation}
\bar{u}_{\bar{r}}(\bar{z}) =\left(\frac{a}{t}\right)\left[\frac{(1-\nu^2)\beta\bar{p}(\bar{z})}{1-(1-\nu^2)\beta\bar{p}(\bar{z})}\right].
\label{eq:Deformed_Radius_Pressure_NonDim__membrane}
\end{equation}
Note that, unlike the case of small-deformation theory leading to Eq.~\eqref{nondimensional_reduced_deflection_microtube}, this last relationship between deformation and pressure is not linear in $p(z)$. Next, substituting $\bar{R}(\bar{z}) = 1 + \left(t/a\right) \bar{u}_{\bar{r}}(\bar{z})$ and Eq.~\eqref{eq:Deformed_Radius_Pressure_NonDim__membrane} into Eq.~\eqref{recipe_for_ODE_numerical} yields an ODE for the pressure $\bar{p}(\bar{z})$. Solving this ODE subject to $\bar{p}(1) = 0$, we obtain
\begin{equation}
\label{eq:PowerLaw_Pressure_Large_Deformation_Dimless}
\bar{p}(\bar{z})=\frac{1}{(1-\nu^2)\beta}\left( 1-\Big\{1+6n(1-\nu^2)\beta[(3+1/n)]^n(1-\bar{z})\Big\}^{-1/3n}\right).
\end{equation}

Next, we evaluate the full pressure drop $\Delta \bar{p}=\bar{p}(0)$ from Eq.~\eqref{eq:PowerLaw_Pressure_Large_Deformation_Dimless} and perform a Taylor series expansion for $\beta\ll1$, to obtain:
\begin{equation}
    \label{eq:large_Deformation_Series}
    \Delta \bar{p}= \Bigg\{2\left(3+1/n\right)^n-\frac{(1+3n)}{2}[(1-\nu^2)\beta]\left[2\left(3+1/n\right)^n\right]^2-\frac{(1+3n)(1+6n)}{6}[(1-\nu^2)\beta]^2\left[2\left(3+1/n\right)^n\right]^3+\mathcal{O}(\beta^3)\Bigg\}.
\end{equation}
On the other hand, a Taylor series expansion of Eq.~\eqref{eq:DP_vs_Q_Microtube} leads to
\begin{equation}
    \label{eq:Small_Deformation_Series}
    \Delta \bar{p}= \Bigg\{2\left(3+1/n\right)^n-\frac{(1+3n)}{2}[(1-\nu^2)\beta]\left[2\left(3+1/n\right)^n\right]^2-\frac{(1+3n)(3+6n)}{6}[(1-\nu^2)\beta]^2\left[2\left(3+1/n\right)^n\right]^3+\mathcal{O}(\beta^3)\Bigg\}.
\end{equation}
Equations~\eqref{eq:large_Deformation_Series} and \eqref{eq:Small_Deformation_Series} agree to \emph{two} terms, with the first discrepancy being the relatively minor change of $1+6n$ becoming $3+6n$ in the coefficient of $\beta^2$. This shows that the pressure drop predicted by the law of Laplace is almost indistinguishable from the corresponding one obtained from small-deformation theory given by Eq.~\eqref{eq:DP_vs_Q_Microtube}.

Now, writing the pressure drop computed from Eq.~\eqref{eq:PowerLaw_Pressure_Large_Deformation_Dimless} in dimensional variables and specializing it to a Newtonian fluid ($n=1$, $m=\mu$), we obtain
\begin{equation}
\label{eq:Newtonian_flow_Rate_Pressure_Large_Deformation}
    \Delta p = \frac{Et}{(1-\nu^2)a}\left\{1-
\left[1+\frac{24\mu q \ell(1-\nu^2)}{\pi a^3Et}\right]^{-1/3}\right\}.
\end{equation}
Equation~\eqref{eq:Newtonian_flow_Rate_Pressure_Large_Deformation} can be compared to the corresponding flow rate--pressure drop relation from Fung's textbook \cite[\S3.4, Eq.~(8)]{F97}:
\begin{equation}
\label{eq:Fung}
\Delta p = \frac{Et}{a}\left\{1-
\left[1+\frac{24\mu q \ell}{\pi a^3Et}\right]^{-1/3}\right\}.
\end{equation}
Clearly, Eqs.~\eqref{eq:Newtonian_flow_Rate_Pressure_Large_Deformation} and \eqref{eq:Fung} are quite similar with Eq.~\eqref{eq:Fung} simply being the case of $\nu=0$ of Eq.~\eqref{eq:Newtonian_flow_Rate_Pressure_Large_Deformation}. The physical reason for this mathematical fact is that, in Fung's analysis, the cylinder's axial stresses and deformation are neglected (yielding a state of \emph{uniaxial} stress), which means that Fung's result is a so-called \emph{independent-ring model}. This approximation is distinct from the shell models considered herein, in which both $\sigma_{\theta\theta}$ and $\sigma_{zz}$ are taken into account. Therefore, Fung's result, being independent of the Poisson ratio, is strictly applicable only to highly compressible solids such as cork, rather than rubber-like elastomers such as PDMS microtubes or blood vessels. 

\begin{figure}
\centering
\subfloat[Newtonian fluid.]{\includegraphics[width=0.4\linewidth]{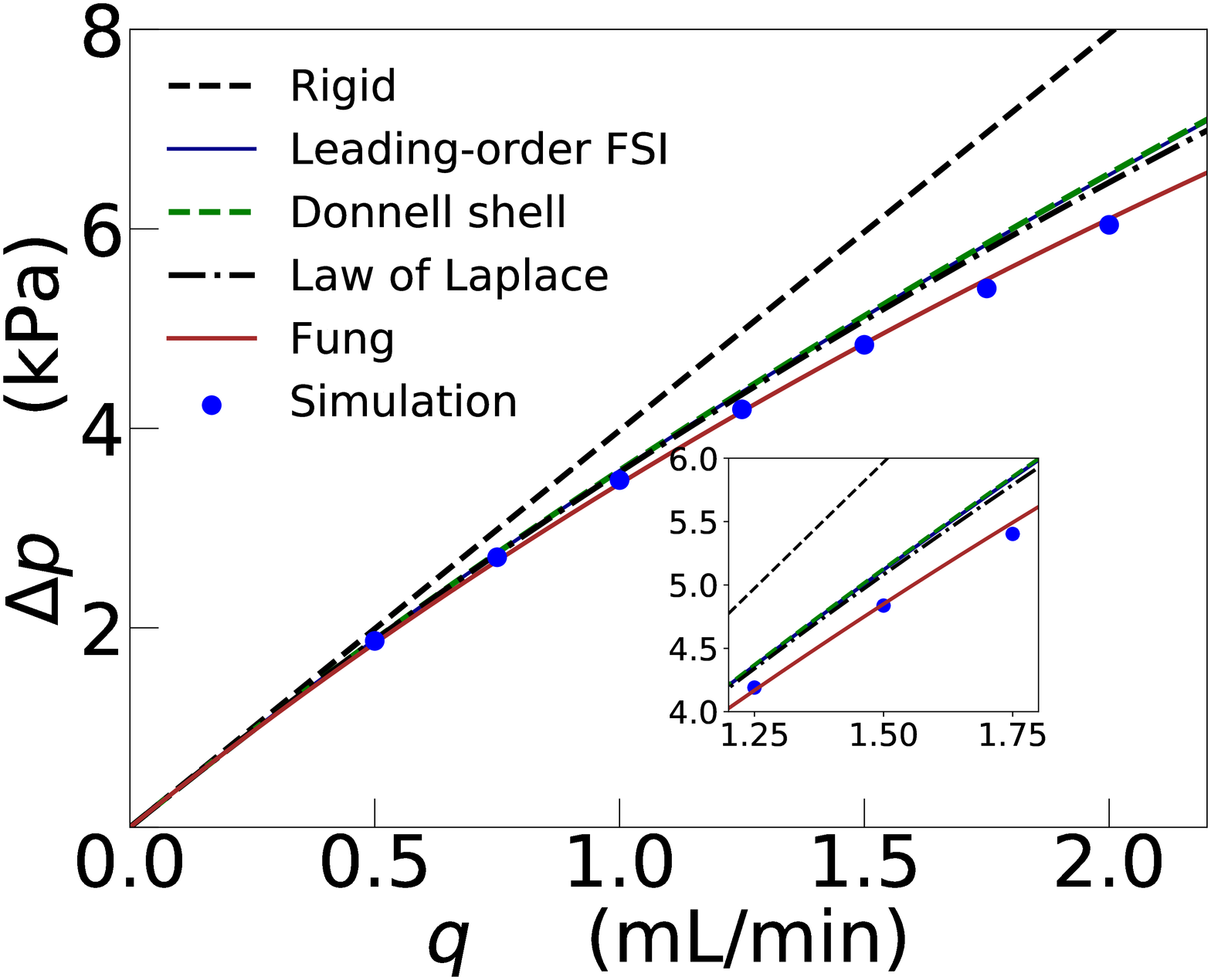}}
\hspace{1.5cm}
\subfloat[Shear-thinning fluid.]{\includegraphics[width=0.4\linewidth]{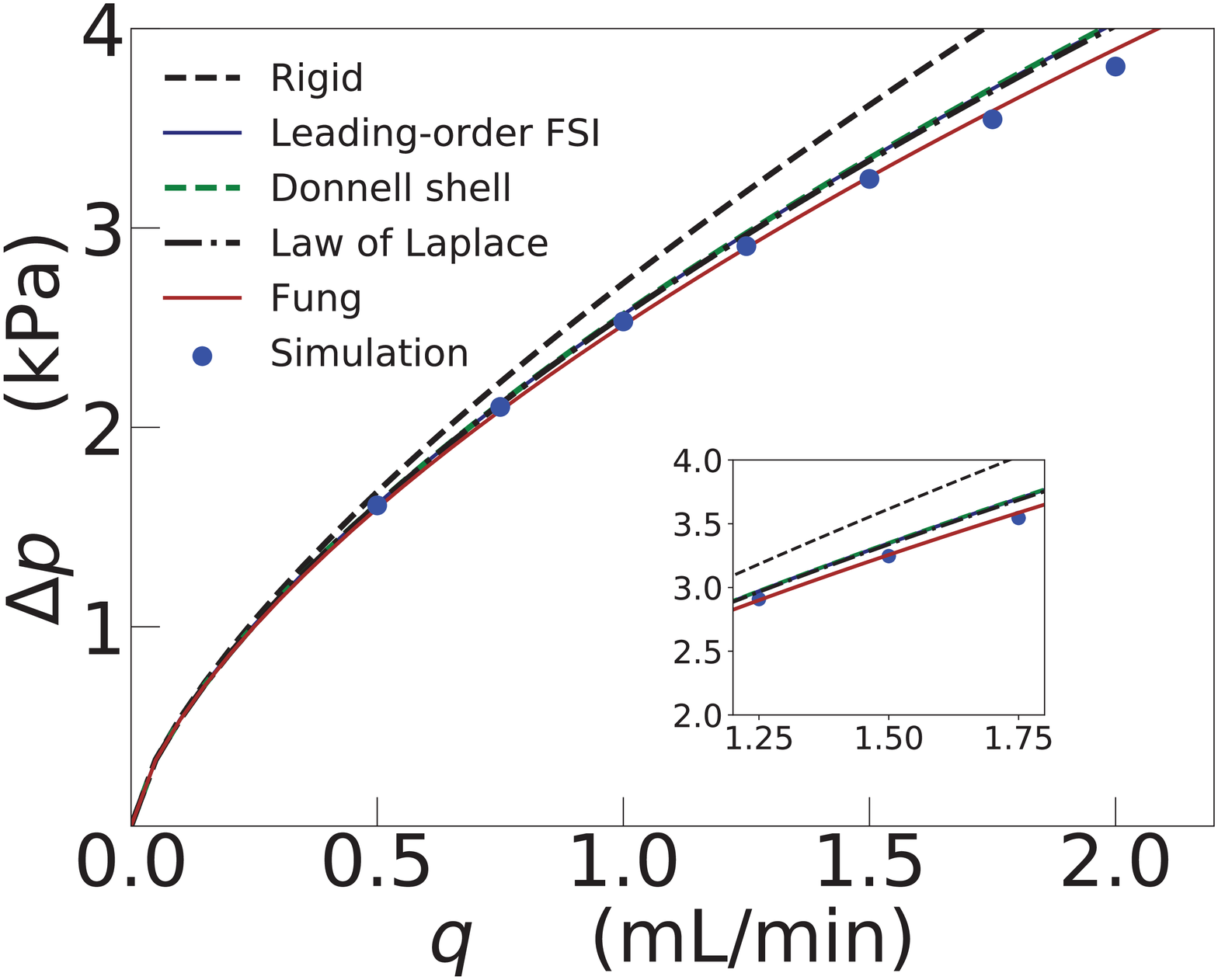}}
\caption{~Flow rate--pressure drop relation for FSI in an elastic tube. The difference between the results from the law of Laplace [large-deformation theory,  Eq.~\eqref{eq:PowerLaw_Pressure_Large_Deformation_Dimless}] and the leading-order FSI theory [small-deformation theory,  Eq.~\eqref{eq:P_vs_Z_Microtube}] is minute. Meanwhile, the results from Fung's relation given by Eq.~\eqref{eq:Fung} deviate significantly from the results of other theories. Note the different the vertical scales in the plots.}
\label{fig:MT_Simulation_Delta_P_Vs_Q_Fung_large}
\end{figure}

\begin{figure}
\subfloat[Newtonian fluid.]{\includegraphics[width=0.5\linewidth]{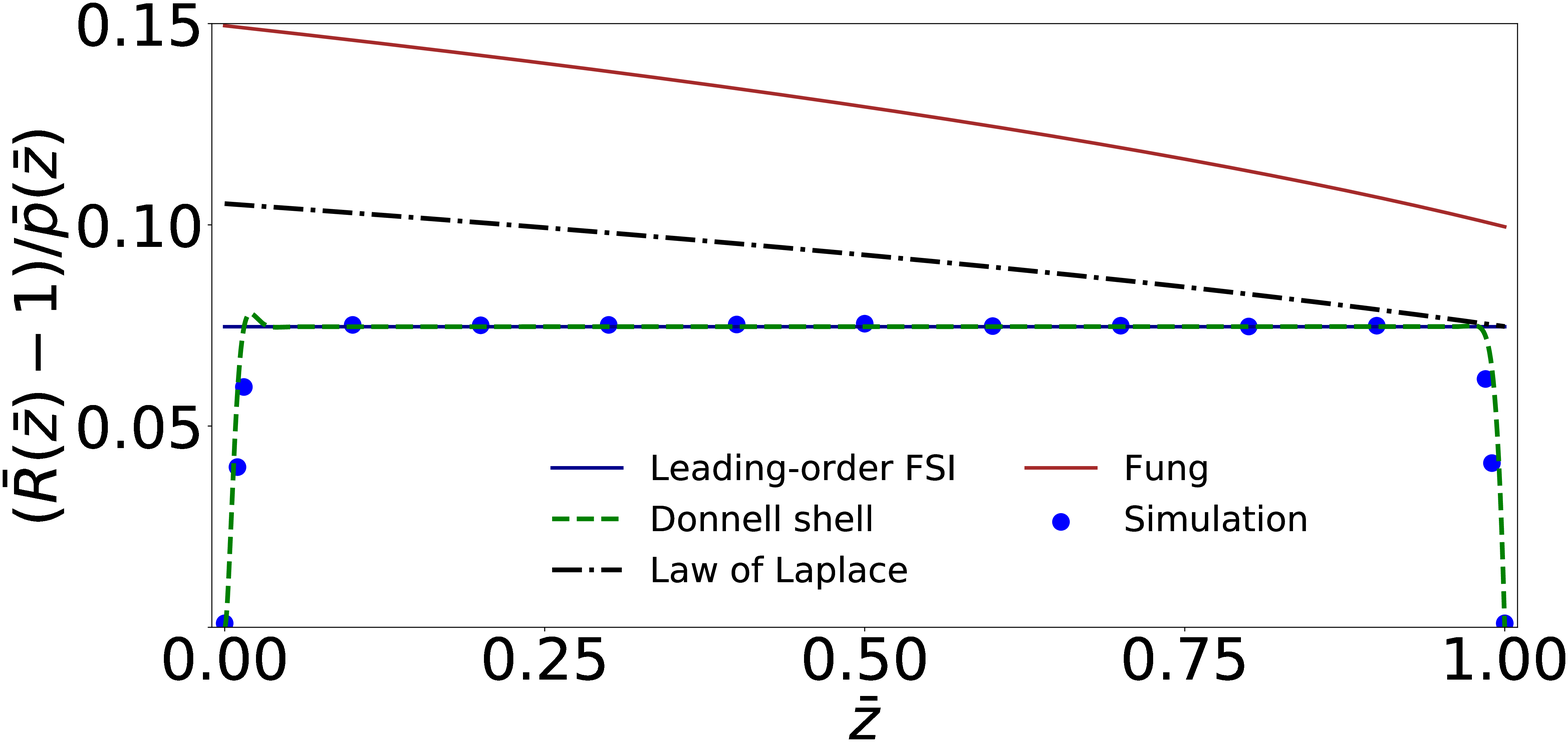}}
\hfill
\subfloat[Shear-thinning fluid.]{\includegraphics[width=0.5\linewidth]{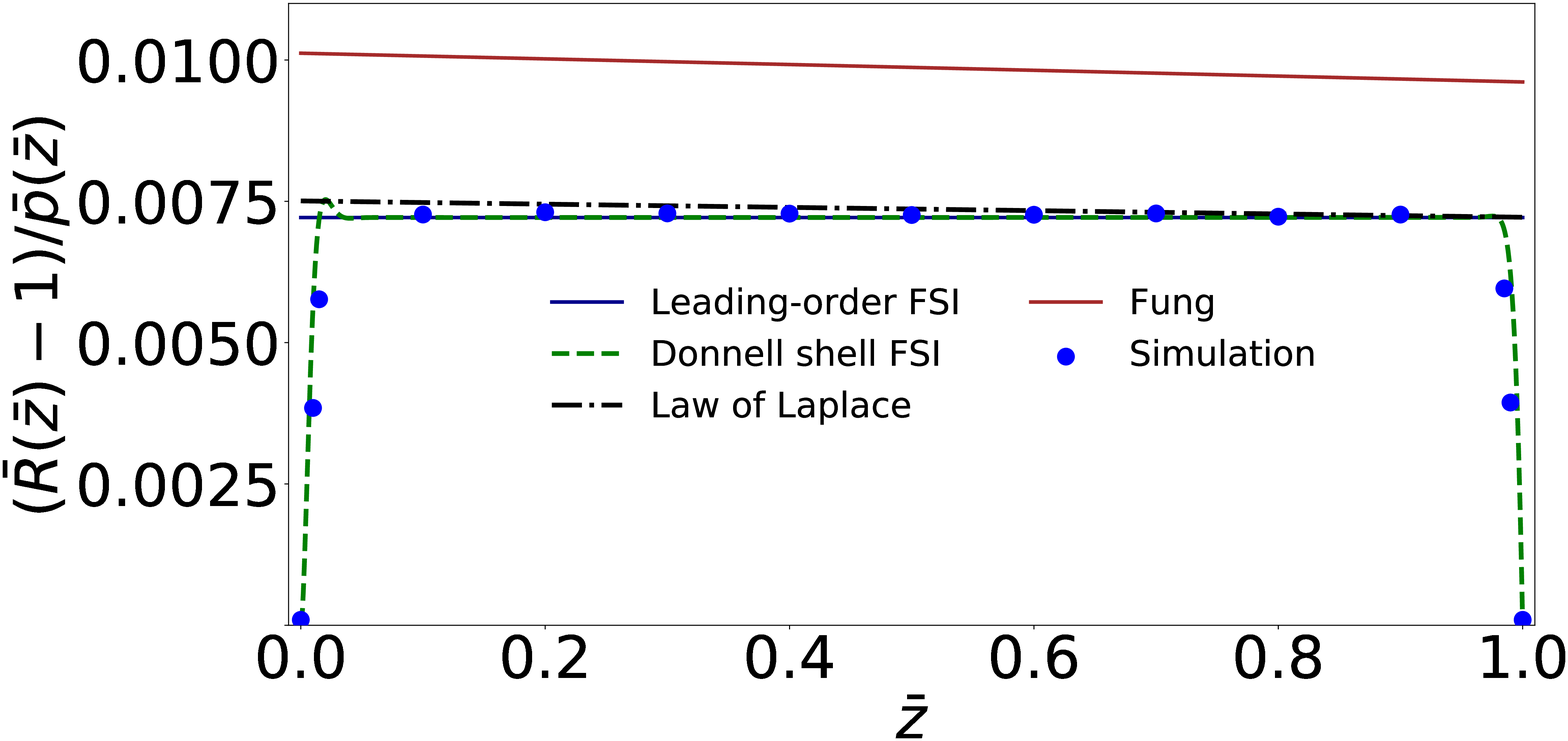}}
\caption{~Ratio of the dimensionless radial deformation $\bar{R}-1$ to the dimensionless hydrodynamic pressure $\bar{p}$, as a function of the dimensionless axial position $\bar{z}$ in the elastic tube for $q = 1$ mL/min. The leading-order FSI theory is given by Eq.~\eqref{nondimensional_reduced_deflection_microtube}, the Donnell shell FSI theory is the numerical solution of the TPBVP given by Eqs.~\eqref{eq:coupled_ODE_non_dim} and \eqref{eq:coupledBC}, the law of Laplace is given by Eq.~\eqref{eq:Deformed_Radius_Pressure_NonDim__membrane}, Fung's result is given by Eq.~\eqref{eq:Deformed_Radius_Pressure_NonDim__membrane} with $\nu=0$, and the simulation results are from ANSYS. Note the different vertical scales in the plots.}
\label{fig:MT_Deflection_Vs_Z_Fung_large}
\end{figure}

In Fig.~\ref{fig:MT_Simulation_Delta_P_Vs_Q_Fung_large}, we compare the results of Fung's model and the law of Laplace, alongside those of our leading-order (membrane) and Donnell-shell FSI theories. The full dimensional pressure drops $\Delta p$ predicted by the different mathematical theories considered are quite close. While Fung's model first overpredicts then underpredicts the pressure drop, the other theories over-predict the pressure drop with increasing error as $q$ increases, consistent with the perturbation approach under which they were derived. As expected, given the small deformation of the tube, the use of a large-deformation membrane theory does not influence the predicted pressure drop. Therefore, we conclude that the assumptions about the deformation and the stress made in Fung's classic treatment of the problem \emph{coincidentally} yield good agreement at large $q$, while Fung's model reduces to the theory derived herein for small $q$.

We also carry out a comparison of the structural mechanics prediction, in Fig.~\ref{fig:MT_Deflection_Vs_Z_Fung_large}, by comparing the ratio $[\bar{R}(\bar{z})-1]/\bar{p}(\bar{z})$ of dimensionless deformation and pressure obtained across the models. Here, the conflicting assumptions used to obtain the law of Laplace and Fung's model are quite apparent in this comparison. For both Newtonian and power-law fluids, Fung's model overpredicts $[\bar{R}(\bar{z})-1]/\bar{p}(\bar{z})$. As explained above, Fung's model is an ``independent ring'' model that neglects the stress and strains in the streamwise direction. These assumptions, in conjunction with those of large-deformation FSI, lead both Fung's model and the law of Laplace to predict an incorrect displacement profile for which $[\bar{R}(\bar{z})-1]/\bar{p}(\bar{z})$ slowly varies with $z$ (outside the boundary and corner layers).

\end{document}